\documentclass[a4paper,12pt]{article}
\usepackage{amsmath}
\usepackage{graphicx}
\usepackage{enumerate}
\usepackage{natbib}
\usepackage{amssymb}
\usepackage{multirow}
\usepackage[caption = false]{subfig} 
\usepackage{comment}
\usepackage{float}
\usepackage{amsthm}
\usepackage{xcolor}
\usepackage{multirow}
\usepackage{subfig}
\usepackage{lscape}

\newcommand{\brackets}[1]{\left(#1\right)}
\newcommand{\E}[1]{\mathbb{E}\left[#1\right]}

\renewcommand{\P}{\mathbb{P}}
\newcommand{\etas}{ETAS}
\newcommand{\retas}{RETAS}

\newcommand{\rh}{RHawkes}
\newcommand{\rmd}{\mathrm{d}}
\newcommand{\set}[1]{\left\{#1\right\}}
\newcommand{\tr}{\mathrm{^\top}}
\newcommand{\norm}[1]{\Vert#1\Vert}

\begin{document}

\title{Stochastic declustering of earthquakes with
	the spatiotemporal \retas\ model}
\author{Tom Stindl\thanks{
		This research includes computations using the Linux computational cluster Katana supported by the Faculty of Science, UNSW Sydney, and the National Computational Infrastructure (NCI) supported by the Australian Government.}  \,\,and Feng Chen\thanks{Chen was partly supported by a UNSW SFRGP grant.} \\
  Department of Statistics, UNSW Sydney} \date{}

\maketitle

\begin{abstract}
  Epidemic-Type Aftershock Sequence (\etas) models are point processes
that have found prominence in seismological modeling.  Its
success has led to the development of a number of different 
versions of the \etas\ model.  Among these extensions is the
\retas\ model which has shown potential to improve the
modeling capabilities of the \etas\ class of models.  The \retas\
model endows the main-shock arrival process with a renewal process
which serves as an alternative to the homogeneous Poisson process.
Model fitting is performed using likelihood-based estimation by
directly optimizing the exact likelihood.  However, inferring the
branching structure from the fitted \retas\ model remains a
challenging task since the declustering algorithm that is 
currently available for the \etas\
model is not directly applicable.
This article solves this
problem by developing an iterative algorithm to calculate the smoothed 
main and aftershock probabilities conditional on all available
information contained in the catalog. Consequently, an objective
estimate of the spatial intensity function can be obtained and an
iterative semi-parametric approach is implemented to estimate model
parameters with information criteria used for tuning the smoothing
parameters. The methods proposed herein are illustrated on
simulated data and a New Zealand earthquake catalog.

\noindent\emph{Key words and phrases:} 	point process, renewal process, 
semi-parametric, self-exciting, seismology
\end{abstract}

\section{Introduction}
\label{sec:Introduction}
Spatiotemporal point processes are useful to model and forecast short
and long-term seismicity. In particular, self-exciting point processes
simultaneously model long-term trends (main-shocks) and
short-term variations (aftershocks) within a unified framework.  The
arrival rate of earthquakes is formed as the superposition of 
two sources of intensity.  The first source is a contribution from the
main-shocks, and the second is an accumulation of the excitation
effects due to past earthquakes which wane over time. These models
have been prominent in modeling the temporal and spatial
clustering of seismic activity in many regions.

The Epidemic-type aftershock sequence (\etas) model was first introduced
in~\cite{Ogata1988} and then later extended to a spatiotemporal model
in~\cite{Ogata1998}. 
The ETAS model has been tailored specifically to model seismicity
based on 
well-researched seismic properties and observed phenomena.
For instance, the temporal response function is based on Omori's law
\citep{Omori1894, Utsu1961} and the distribution of magnitudes based
on Gutenberg-Richter's law \citep{Gutenberg1944}. Since the model's inception, 
there have been many modifications to enhance its  modeling 
capabilities 
~\citep{Zhuang2002, Console2003, Ogata2011, Guo2015, Fox2016a, Cheng2018, Stindl2021}. 
Not only has the
\etas\ model 
been successful in modeling earthquake catalogs but it has also shown
strong potential to model other phenomena such as crime
\citep{Mohler2011, Mohler2014, Zhuang2019}, spread of diseases
\citep{Meyer2012, Schoenberg2019}, social networks \citep{Fox2016,
	Zipkin2016}, wildfires \citep{Peng2005} and terrorist activity
\citep{Clark2018}.

The \etas\ model has been the cornerstone for seismological
modeling. However, there exists some debate on the validity of the
homogeneous Poisson process assumption for the occurrence times of
main-shocks. For instance, Stress-Release
theory 
suggests that the main-shock arrival process should be time-dependent
such as in the form of a renewal
process. 
Reid's elastic rebound theory~\citep{Reid1910} suggests that
earthquakes occur as a result of the release of energy from the
accumulation of strain energy along faults. For consistency with
Reid's theory, the intensity for main-shocks
should depend on the time of the most recent main-shock, that is, the
time in which energy was last
released. 

The renewal \etas\ (\retas) model is a 
spatiotemporal point process 
proposed by~\cite{Stindl2021} motivated by Reid's Stress-Release
theory. The \retas\
model 
introduces heterogeneity in the background rate by specifying a
renewal arrival process for main-shocks whereby the intensity for
main-shocks resets at the main-shock arrival times. The \retas\ model
is an extension of the renewal Hawkes (RHawkes) process proposed
by~\cite{Wheatley2016} but encompasses a spatial component to
accommodate the spatial clustering of earthquakes and 
tailored parametric forms for the excitation effects. Estimation of model
parameters for the \retas\ model is performed using maximum likelihood
(ML) based on an iterative algorithm for likelihood evaluation similar
to that for the \rh\ process
\citep{Chen2018}
. 

\etas\ model fitting requires the estimation of a spatial 
intensity function which plays a crucial in both model 
fitting and forecasting. The spatial intensity function identifies 
regions with persistent and strong incidence of seismic activity, 
independent of the aftershock clustering features which wane over time. 
It is common to estimate the spatial intensity function
by utilising a stochastic declustering algorithm. 
For each earthquake, the
delcustering algorithm provides an estimated main-shock and aftershock
probability for each possible triggering earthquake.
The main-shock probabilities are used as weights to fit a weighted 2-d
kernel density estimator (KDE) for the spatial intensity function. The
estimates of the other parameters of the \etas\ model are then 
updated to accommodate the most recent estimate of the spatial intensity
function. Then, the main-shocks weights will also need to adjust, and
re-estimation of the spatial intensity function is required.
These two steps are repeated until convergence. 

However, estimation of the spatial intensity function
for the \retas\ model has been confined to using historical data, and no 
objective estimation technique based on the observed catalog
is currently available. This is because the declustering algorithm for the 
\etas\ model
\citep{Zhuang2002} is not applicable in the case of the \retas\ model due to the
intricate dependence of its event intensity on past 
earthquakes. 
Therefore, when performing stochastic declustering for the \retas\
model, the main and aftershock probabilities must be calculated by
conditioning on the complete observed catalog including both past and
future earthquakes. That is, the smoothed probabilities are needed,
rather than the filtered probabilities which the \etas\ model only requires. 

This article develops a declustering algorithm that accounts for the
dependence structure of the \retas\ model. To this end, we propose a
backward smoothing procedure to calculate the most recent 
main-shock probabilities but conditioned on the complete observed 
catalog. 
These probabilities facilitate the calculation of main and aftershock 
probabilities which are in a form amenable to stochastic declustering 
for the \retas\ model. 
As a consequence, an objective estimate of the spatial intensity
function of the \retas\ model based on the observed catalog can be
obtained without resorting to the use of historical data or other
simplistic assumptions. Fitting the \retas\ model to earthquake
catalogs then proceeds similar to the \etas\ iterative semi-parametric 
procedure. 

This article also proposes a data-driven procedure to select an
appropriate amount of smoothing to be used in estimating the spatial intensity
function. The procedure is based on the corrected Akaike information
criterion (AICc), which requires the effective number of parameters of the
smoothed spatial intensity function calculated for the fixed smoothing
matrix.  This effective number of parameters does not change over the
estimation iterations, unlike the weights which need to be updated for
different parameter estimates. After the final iteration of the
estimation algorithm the corrected AIC can be calculated to compare
among different choices of the smoothing matrix. We show using
simulations that this procedure provides a suitable strategy for smoothing
parameter selection and leads to parameter estimates that are
comparable with the ML estimates obtained assuming the true parametric
form of the spatial intensity
function
.

The rest of this article contains the following.
Section~\ref{sec:RETAS-Model} details the general form of the
\retas\ model and outlines the iterative log-likelihood evaluation
algorithm. Section~\ref{sec:declust-alg} describes a stochastic
declustering algorithm for the \retas\
model. 
Section~\ref{sec:simulations} reports the result of our numerical
experiments to investigate the declustering algorithm and the
iterative semi-parametric estimation procedure using simulated earthquake
catalogs. This section also includes a comparison to the declustering 
algorithm of the \etas\ model when (inappropriately) applied to the \retas\ model.
Section~\ref{sec:applications} analyzes an earthquake catalog
from New Zealand (NZ) by applying the proposed methodologies.

\section{\retas\ model and likelihood evaluation}
\label{sec:RETAS-Model}
Let $\set{(t_i,x_i,y_i,m_i),\ i=1,\dotsc,n}$ be an earthquake
catalog, where we denote by $t_i \leq T$ the occurrence time,
$(x_i,y_i)\in \mathcal{S}$ the coordinates of the epicentre and
$m_i \geq m_0$ the magnitude of the $i$th earthquake, with $T \in \mathbb{R}_+$ being
the censoring time, $\mathcal{S}\subset \mathbb R^2$ the spatial
region, and $m_0$ the threshold magnitude of the catalog. Let
$N(\cdot)$ denote the point process associated with the catalog
and $N(A^t\times A^x\times A^y\times A^m)$ denote the number of
earthquakes with occurrence times in $A^t$, epicentres in
$A^x \times A^y$, and magnitudes in $A^m$.
The conditional intensity function of $N$ is defined as
\begin{multline}
\label{eq:intensity-def}
\lambda(t,x,y,m| \mathcal{H}_{t-}):= \\
\lim\limits_{\Delta t, \Delta x,
	\Delta y, \Delta m \rightarrow 0}
\frac{\E{N \brackets{[t,t+\Delta t) \times [x,x+\Delta x) \times
			[y,y+\Delta y) \times[m,m+\Delta m)}  | 
		\mathcal{H}_{t-}}}{\Delta t \Delta x \Delta y \Delta m},
\end{multline} 
where
$\mathcal{H}_{t-} = \set{(t_i,x_i,y_i,m_i) \, ; \, t_i <
	t}$ denotes the pre-$t$ history and represents the complete knowledge 
of times, locations and magnitudes of earthquakes up to but not including 
time $t$.

The \etas\ model \citep{Ogata1998} assumes that the conditional intensity
in~\eqref{eq:intensity-def} takes the separable form
\begin{equation}
\lambda(t,x,y,m | \mathcal{H}_{t-}) = \lambda(t,x,y | \mathcal{H}_{t-})  J(m),
\end{equation}
where the density function $J(m)$ is independent of all other model
components, 
and the intensity $\lambda(t,x,y | \mathcal{H}_{t-}) $ is
self-exciting and given by
\begin{equation}
\lambda(t,x,y | \mathcal{H}_{t-}) = \mu_0 \, \nu(x,y) +
\sum_{i:\,t_i<t} \psi(t-t_i,x-x_i,y-y_i;m_i), 
\end{equation}
where $\mu_0$ is a constant temporal main-shock arrival rate,
$\nu(x,y)$ is the spatial intensity function which distributes 
the main-shocks in $\mathcal{S}$ and $\psi(t,x,y;m)$ is the triggering
function for aftershocks. The \etas\ model further refines the triggering function
$\psi$ into three separate functions $g(t)$, $f(x,y)$ and $k(m)$
pertaining to time, space and magnitude of the aftershocks
respectively
\begin{equation}
\psi(t-t_i,x-x_i,y-y_i;m_i) = g(t-t_i)f(x-x_i,y-y_i)k(m_i).
\end{equation}
The temporal response function $g(t)$ and spatial response function
$f(x,y)$ describe how the conditional rate of earthquakes decay over
time and space, respectively, while the boost function $k(m)$
measures the influence of magnitudes from past earthquake, 
in which larger magnitude earthquakes are more productive at 
triggering additional aftershocks than smaller 
ones.

The \retas\ model replaces the constant temporal main-shock rate
$\mu_0$ with a time-dependent function that renews 
at the arrival time of main-shocks. 
To introduce the \retas\ model, we assume an earthquake is either a
main-shock or an aftershock induced by any previous earthquake.
Write $B_i = j$ $(<i)$ if the $i$th earthquake was induced by the $j$th
earthquake, otherwise $B_i = 0$ if the $i$th earthquake is a
main-shock. Then $I(t) = \max\set{i; t_i < t, B_i = 0}$ represents
the (unobserved) index of the last main-shock prior to time $t$. The
\retas\ model assumes the conditional intensity function takes
the form
\begin{align}
\lambda(t,x,y | \tilde{\mathcal{H}}_{t-})
& = \mu(t-t_{I(t)}) \,
\nu(x,y) + \sum_{i:
	t_i<t}
g(t-t_i)f(x-x_i,y-y_i)k(m_i),  
\label{eq:retas-intensity} \\
& = \mu(t-t_{I(t)}) \, \nu(x,y) + \phi(t,x,y), \notag
\end{align}
where $\tilde{\mathcal{H}}_{t-} = \mathcal{H}_{t-} \cup \set{I(t)}$ is
the augmented information set which encompasses the pre-$t$ history
and the index of the last main-shock prior to time $t$. The intensity 
in~\eqref{eq:retas-intensity} is defined with respect to the extended 
history $\tilde{\mathcal{H}}_{t-}$ and not the pre-$t$ history 
$\mathcal{H}_{t-}$. The reason for this is because the intensity 
with respect to the pre-$t$ history has a complex form due to the 
intricate dependence of the event intensity on past points while the 
intensity conditioned on the extended history has a convenient form for 
presenting the \retas\ model and its log-likelihood function.

The following parametrization will be used throughout the article. The
waiting times between main-shock arrivals are independent and gamma
distributed with hazard rate function
\begin{equation}
\label{eq:hazGam}
\mu(t) = \frac{1}{\Gamma(t/\beta,\kappa) \beta^\kappa}
t^{\kappa-1} e^{-t/\beta}, \quad t >  0, 
\end{equation}
where $\kappa >0$ is the shape parameter, $\beta>0$ is the scale parameter
and $\Gamma(x,k)=\int_x^\infty s^{k-1} e^{-s}\rmd s$ is the upper
incomplete gamma function. The temporal response function $g$ is derived from the 
modified Omori's law \citep{Utsu1961} and takes the form
\begin{equation}
\label{eq:omori-law}
g(t) = \frac{p-1}c\bigg(1+\frac t c\bigg)^{-p}, \quad t >  0,
\end{equation}
where $p > 1$ is a shape parameter indicating the rate of aftershock decay and $c>0$ is a scale parameter. The spatial response function $f$ is 
bivariate normal with independent marginals 
\begin{equation}
\label{eq:biv-norm}
f(x,y) = \frac{1}{2 \pi \sqrt{\sigma_1^2 \sigma_2^2}}\exp \brackets{-\frac{x^2}{2\sigma_1^2} - \frac{y^2}{2\sigma_2^2}}, \quad (x,y) \in \mathbb{R}^2,
\end{equation}
where $\sigma_1^2>0$ and $\sigma_2^2>0$ are the variances in the $x$- and $y$-directions respectively. 
The boost function takes the exponential form
\begin{equation}
\label{eq:k-fn}
k(m) = Ae^{\alpha(m - m_0)}, \quad m \ge m_0,
\end{equation}
where 
$A > 0$ controls the average number of induced aftershocks and
$\alpha \in \mathbb{R}$ reflects the relative influence of the
magnitudes on the intensity. Since both $g$ and $f$ are density
functions, the boost function $k(m)$ indicates the expected count
of aftershocks induced by a magnitude $m$  earthquake.  The
distribution of the magnitudes is motivated by the Gutenberg-Richter
law \citep{Gutenberg1944} and follows a shifted exponential
distribution with density
\begin{equation}
\label{eq:shifted-exp}
J(m) = \gamma e^{-\gamma(m-m_0)}, \quad m\geq m_0,
\end{equation}
where $\gamma >0$ is a scale parameter. Under this 
parameter formulation, the productivity (Prod.) 
is the expected count of aftershocks induced by 
a single earthquake and is given by 
$
\mathbb{E} \big[Ae^{\alpha(m_i - m_0)}\big] = A\gamma
/(\gamma-\alpha),
$
which must be less than one to guarantee stationarity.

When the spatial intensity function $\nu(x,y)$ is assumed known, or at
least fixed at some estimate, we can estimate 
the parameters 
$\theta = (\kappa,\beta,p,c,\sigma_1^2,\sigma_2^2$, $A,\alpha)$ by directly
optimizing the log-likelihood function, which 
can be evaluated using the recursive algorithm proposed in
\cite{Stindl2021}. 
For estimation of the spatial intensity function $\nu(x,y)$,
non-parametric techniques are generally required. Strategies include
2-d weighted KDEs \citep{Musmeci1992, Zhuang2002}, or bi-cubic
B-splines based on the identified main-shocks obtained from a 
deterministic magnitude
based declustering algorithm 
\citep{Ogata1998}. This article employs a 2-d weighted KDE and
utilizes the estimated smoothed main-shock probabilities from the
stochastic declustering algorithm discussed herein to 
obtain an objective estimate.

For ease of reference, we reproduce the log-likelihood evaluation
algorithm provided in \cite{Stindl2021}. 
Let the filtered probabilities for the index of the last main-shock be 
denoted by $p_{ij}=\P(I(t_i) = j|\mathcal{H}_{t_i -})$, and
further define
$d_1 = \mu(t_1) \nu(x_1, y_1) e^{-\int_{0}^{t_1}\mu(t) \rmd t
}$, $S_{n+1,j} = e^{-\int_{t_n}^{T} \mu(t - t_j) \rmd t}$ and
for $i =2,\dotsc,n$,  \(j=1,\dotsc,i-1\), let
\begin{align*}  
d_{ij} & = \brackets{\mu(t_i - t_j)\nu(x_i,y_i) + \phi(t_i,x_i,y_i) } 
S_{ij},\\
S_{ij}&=e^{-\int_{t_{i-1}}^{t_i} \mu(t-t_j)  \rmd t},
\end{align*} 
and
\begin{equation*}
\Phi(t)= \sum_{j:t_j<t} k(m_j) \int_{t_j}^t g(s - t_j)\,\rmd
s \iint_{\mathcal{S}}f(x - x_j,y-y_j)\,\rmd x\rmd y.
\end{equation*}
The log-likelihood of the \retas\ model is given by
\begin{multline}
\label{eq:log-lik}
\ell(\theta ) = \log d_1 + \sum_{i=2}^{n} 
\log \brackets{ \sum_{j=1}^{i-1}p_{ij}\,d_{ij} } + 
\log \brackets{\sum_{j=1}^{n} p_{n+1,j} S_{n+1,j}  } + \Phi(T) + 
\ell_m,
\end{multline}
where $\ell_m = \sum_{i=1}^{n} \log J(m_i)$ and the $p_{ij}$'s are 
calculated recursively by
\begin{equation}
\label{eq:pij-rec}
p_{ij}  = \begin{cases} \displaystyle
\frac{
	p_{i-1,j} \phi(t_{i-1},x_{i-1},y_{i-1}) S_{i - 1,j}}{
	\sum_{j=1}^{i-2} p_{i-1,j}d_{i-1,j}}, & j = 1,\dotsc,i - 2 \\
1-\sum_{k = 1}^{i - 2}p_{ik}, & j = i - 1
\end{cases},\quad i\geq 3,
\end{equation}
with the initial condition $p_{21}=1$.

\section{Stochastic Declustering algorithm}
\label{sec:declust-alg}
This section provides a declustering algorithm for the \retas\ model
that calculates the smoothed main-shock and triggering (parent) probabilities
for each earthquake in the catalog. This allows us to infer which
earthquakes are main-shocks or aftershocks 
and hence derive an objective estimate of the spatial intensity
function $\nu(x,y)$ based only on the observed earthquake catalog. 
It also facilitates inferences to be drawn from the catalog 
such as highly productive main-shocks leading to large clusters and direct links 
between earthquakes.

\subsection{Smoothed probabilities for stochastic declustering}

The objective of declustering is to estimate the main and aftershock probabilities for each earthquake in the catalog. 
Let $\pi_{ij}= \P \brackets{B_i = j \mid \mathcal{H}_T}$ denote the
conditional probability that the $j$th earthquake triggers the $i$th
earthquake 
conditional on the complete information set
$\mathcal{H}_T=\{t_{1:n},x_{1:n},y_{1:n},m_{1:n},$ $t_{n+1}>T\}$ where here
and hereafter, we use the shorthand notation
$v_{j:i}=(v_j,v_{j+1},\dotsc,v_{i})$ for $j < i$. Let
$\omega_{ij} = \P \brackets{B_i=0, I(t_i)=j \mid \mathcal{H}_T}$ be
the probability that the $i$th earthquake is a main-shock and the last
main-shock prior to it has index $j$, and
$\omega_i = \P(B_i = 0 | \mathcal{F}_T) =\sum_{j=1}^{i-1} \omega_{ij}$ denotes 
the smoothed probability that the $i$th earthquake is a main-shock.
Let $q_{ij} = \mathbb{P}\left(I(t_i) = j \mid\mathcal{H}_T \right)$
denote the smoothed probability that the $j$th earthquake is the last
main-shock before the $i$th earthquake.    

The probabilities derived from the declustering algorithm are based on
the distribution of $B_{1:n} 
$ conditional on the complete information set $\mathcal{H}_T$.  For
the standard \etas\ model, $B_{1:i}$ is conditionally independent of
$\{t_{i+1:n}, x_{i+1:n}$, $y_{i+1:n}, m_{i+1:n}, t_{n+1}>T
\}$ given
$\mathcal{H}_{t_i}$, 
which is not generally true of the \retas\ model. For correct
application of the declustering algorithm, the full conditional
distribution of $B_{1:n}$ given $\mathcal{H}_T$ is required, which can be
obtained by implementing a backward recursion starting with the
filtered probabilities $p_{n+1, j}$ calculated during likelihood evaluation 
in~\eqref{eq:pij-rec}. 

The smoothed most recent main-shock probabilities for $i = n,\dotsc,2$ are given by
\begin{equation}
q_{ij} = \frac{f_{ij}}{\sum_{k=1}^{i-1} f_{ik} p_{ik}} \, p_{ij}, \qquad j = 1,\dotsc,i-1,
\label{eq:qij-rec}
\end{equation}
where $f_{ij} = p(t_{i:n}, x_{i:n}, y_{i:n}, m_{i:n}, t_{n+1} > T | \mathcal{H}_{t_i-}, I(t_i) = j)$ are computed recursively by
\begin{equation}
f_{ij} = e^{-\int_{t_{i-1}}^{t_{i}} \mu(s-t_j)\rmd s  -(\Phi(t_{i})-\Phi(t_{i-1}))} \bigg[f_{i+1,j}
\, \phi(t_i,x_i,y_i,m_i)+ f_{i+1,i}
\, 	\mu(t_i - t_j)\nu(x_i,y_i)\bigg], 
\label{eq:fij-rec}
\end{equation}
with initial conditions
$f_{n+1,j} = e^{-\int_{t_{n}}^{T} \mu(s-t_j)\rmd s - \{\Phi(T) - \Phi(t_n)\}}$ for $j = 1,\dotsc,n$. The derivation of~\eqref{eq:qij-rec} and~\eqref{eq:fij-rec} can be found in Appendix~\ref{appA}.
With the smoothed most recent main-shock probabilities $q_{ij}$ available they facilitate the calculation of the branching structure probabilities $\omega_i$ and $\pi_{ij}$.
The smoothed main-shock probabilities for $i=n,\dotsc,2$ are given by 
\begin{equation}
\omega_{i} = \sum_{j=1}^{i-1} \omega_{ij},
\label{pi-rec}
\end{equation}
where 
\begin{equation}
\omega_{ij} = \frac{f_{i+1,i} \, \mu(t_i - t_j)\nu(x_i,y_i)}{f_{i+1,i} \, \mu(t_i - t_j)\nu(x_i, y_i) + f_{i+1,j} \, \phi(t_i,x_i,y_i,m_i)} \, q_{ij}, 
\label{eq:omega-rec}
\end{equation}
for $j = 1,\dotsc,i-1$.
%
%
The triggering (parent) aftershock probabilities for $i = n,\dotsc, 2$ are given by
\begin{equation}
\pi_{ij} = \sum_{k=1}^{i-1} \frac{f_{i+1,k} \, k(m_i)  g(t_i - t_j) f(x_i - x_j,y_i - y_j)}{f_{i+1,i} \, \mu(t_i - t_k)\nu(x_i,y_i) + f_{i+1,k} \, \phi(t_i, x_i, y_i, m_i)} \, q_{ik},
\label{eq:pi-rec}
\end{equation}
for $j = 1,\dotsc,i-1$.
The derivation of~\eqref{eq:omega-rec} and~\eqref{eq:pi-rec} can be found in Appendix~\ref{appB}.
Therefore, to compute the smoothed branching structure probabilities we perform for $i = n,\dotsc,2$ the following steps once the $p_{ij}$ in~\eqref{pi-rec} have been calculated during ML estimation; apply the backward recursion in~\eqref{eq:fij-rec} to obtain $f_{ij}$, compute $q_{ij}$ in~\eqref{eq:qij-rec}, then compute the smoothed branching structure probabilities $\pi_{ij}$ and $\omega_{ij}$ using~\eqref{eq:omega-rec} and~\eqref{eq:pi-rec} and then the main-shock probabilities $\omega_i$ in~\eqref{pi-rec}. 

\subsection{Filtered probabilities for stochastic declustering}
\label{sec:filtered}
This section outlines the declustering algorithm that would be obtained 
if we incorrectly used the filtering probabilities in the same way as the 
declustering algorithm of the \etas\ model. The filtered main-shock probabilities 
would be given by
\begin{equation}
\label{eq:omega-etas}
\omega_i^f = \P(B_i = 0 | \mathcal{H}_{\tau_i-}) = \sum_{k=1}^{i-1} \frac{\mu(t_i - t_k)}{\mu(t_i - t_k) + \phi(t_i,x_i,y_i)} p_{ik},
\end{equation}
and the triggering or parent probabilities are given by
\begin{equation}
\label{eq:pij-etas}
\pi_{ij}^f = \P(B_i = j |  \mathcal{H}_{\tau_i-}) = \sum_{k=1}^{i-1} \frac{k(m_i) g(t_i - t_j)f(x_i-x_j,y_i-y_j)}{\mu(t_i - t_k) + \phi(t_i,x_i,y_i)} p_{ik},
\end{equation}
where the superscript $f$ indicates that these are the filtered probabilities. 

\subsection{Estimation of the spatial intensity function}

Stochastic declustering facilitates an objective estimation procedure
for the spatial intensity function
$\nu(x,y)$. 
For the weighted 2-d KDE, each earthquake  in the catalog is
provided a weight that corresponds to its estimated smoothed
main-shock probability $\hat\omega_i$, that is
\begin{align*}
\hat \nu(x,y)=\sum_{i=1}^{N(T)}|\det h|^{-1/2}K(\norm{
	h^{-1/2}(x_i-x,y_i-y)\tr }) \hat\omega_i\bigg/\sum_{j=1}^{N(T)}\hat\omega_j,
\end{align*}
where $K(\cdot)$ is a kernel function, such as the standard normal
density function; $\Vert\cdot\Vert$ denotes the
Euclidean norm; and $h$ the smoothing parameter matrix.

Since there is an interplay between the parameters of the \retas\
model and the branching structure probabilities, it is necessary to
update these quantities until they stabilize.
This leads to the following semi-parametric recursive algorithm for
fitting the \retas\ model:
\begin{enumerate}
	\item 
	Obtain an initial estimate of the spatial intensity function
	$\hat\nu(x,y)$ using a 2-d KDE with equal weights given to all the
	earthquakes in the catalog; that is $\hat\omega_i\equiv 1$. 
	\item Estimate the parameter vector
	$\theta = (\kappa,\beta,p,c,\sigma_1^2,\sigma_2^2,A,\alpha)$ of the
	\retas\ intensity by directly maximizing the log-likelihood function
	in~\eqref{eq:log-lik} with the most recent update for
	$\hat{\nu}(x,y)$ treated as fixed during the optimization routine.
	\item Calculate the smoothed main-shock probabilities $\hat\omega_i$ for
	$i=1,\dotsc,n$ based on the current estimates $\hat\theta$ and
	$\hat{\nu}(x,y)$ using the declustering algorithm.
	\item Update the estimate of the spatial intensity function using the
	weighted 2-d KDE with weights equal to most recent update for
	$\hat\omega_i$.
	\item If the convergence criterion is achieved, stop. Otherwise, return
	to Step~2.
\end{enumerate}
One possible convergence criterion is that the change in the
log-likelihood function in successive iterations is below a small
threshold value.

\subsection{Determining the smoothing parameter of the KDE}
\label{sec:smoothing}
The question then arises as to the amount of smoothing to be used for
the weighted 2d-KDE. We suggest a data-driven approach for selecting the
level of smoothing by minimizing the AICc \citep[corrected Akaike
information criterion;][]{Hurvich1989}, which is a bias-corrected form
of the AIC \citep[Akaike information criterion;][]{Akaike1971} given
by 
$$
\text{AIC}_c = \text{AIC} + 2\frac{k(k+1)}{n - k - 1} =
-2\ell(\hat\theta) + 2k + 2\frac{k(k+1)}{n - k - 1}=-2\ell(\hat\theta)
+ 2\frac{nk}{n - k - 1},
$$
where $\ell(\hat\theta)$ is the maximized log-likelihood, $k$ is the
(effective) number of parameters and $n$ is the sample size. To determine $k$, we need the
effective number of parameters in the KDE $\hat\nu(\cdot,\cdot)$. For
this purpose, we follow the procedure proposed recently
by~\cite{McCloud2020}, 
which expresses the KDE as a linear smoother and takes the trace of
the hat matrix as the effective number of parameters 
\begin{equation}
\text{DoF}
= \sum_{i=1}^n
\frac{\mathcal{W}_h((x_i,y_i),(x_i,y_i))}{\sum_{l=1}^n
	\mathcal{W}_h((x_l,y_l),(x_i,y_i))}, 
\end{equation}
where $ \mathcal{W}_h$ is given by
\begin{align}
\mathcal{W}_{h}((x_j,y_j),(x_i,y_i))=|\det h|^{-1/2}
K(\Vert h^{-1/2}(x_i-x_j,y_i-y_j)\tr\Vert ). 
\end{align}
We propose the AICc rather than the AIC, since the AICc is
preferred to the AIC in situations in which the sample size $n$ is
small or when the number of parameters $k$ is large relative to the
sample size \citep{Burnham2002}. Since the effective number of
parameters for the KDE is bounded above by the sample size, the number
of parameters can be large relative to $n$ and hence the correction
term in the AICc is significant. Based on 
a finite number of
different choices of the smoothing parameter matrix $h$, we perform the
semi-parametric estimation procedure and calculate the AICc value, and
finally select the $h$ that has the smallest AICc value.

\section{Simulations}
\label{sec:simulations}
This section evaluates the numerical performance of the stochastic
declustering algorithm to infer the true branching structure, and the
semi-parametric estimation method to objectively estimate the spatial
intensity function $\nu(x,y)$ and recover the \retas\ model
parameters. 
A comparison of the \retas\ declustering algorithm proposed herein and the 
\etas\ declustering algorithm is presented. We also provide a comparison of 
the \etas\ and \retas\ models accuracy in correctly inferring the true 
branching structure.

\subsection{Simulation model}
\label{sec:sim-mod}
The simulation model used in this section is identical to the model
described in Section~\ref{sec:RETAS-Model} with the following
parameter choices: $\kappa = 0.8$, which is typical for earthquake
catalogs with stronger clustering than a Poisson
process; 
$\beta = 1.25$, which implies a mean waiting time between
main-shocks 
equal to one; $p=1.2$ and $c=0.01$, which is in line with the fitted 
values in the real-data example discussed later; 
$\sigma_1^2 = 0.01$ and
$\sigma_2^2 = 0.02$, which implies that aftershocks are distributed with
twice the variance in the $x$-direction than the $y$-direction;
$A=0.5$, \(\alpha=1\), and \(\gamma=5\), which implies that an
earthquake directly triggers 0.625 earthquakes on
average. 
The spatial region is the whole 2d-space
$\mathcal{S} = \mathbb{R}^2$ and the spatial intensity function
$\nu(x,y)$ is bivariate normal with independent marginals with variances
0.05 and 0.10 in the $x$-direction and $y$-direction respectively.

\subsection{Estimation results}
\label{sec:sim-est}
We simulate $N=1,\!000$ earthquake catalogs from the simulation
model with two different censoring times $T=250$ and $T=500$. 
For each catalog, we estimate the model parameters under two different 
scenarios. First, we assume the correct form for the 
parametric spatial intensity function, and in the second scenario we assume 
it is unknown and use the semi-parametric iterative procedure as
discussed in Section~\ref{sec:declust-alg} with different amounts of
smoothing for the weighted 2d-KDE. The default smoothing matrix 
is selected based on the multivariate plug-in bandwidth selection procedure of~\cite{Wand1994}, as implemented in the \texttt{ks} package in 
\texttt{R}. Different amounts of smoothing were achieved by multiplying 
the default bandwidth matrix by a factor $\zeta$ in the set $\set{0.5,1,1.5,2,2.5,3}$.

The estimation results under the scenario when the true spatial intensity
function $\nu(x,y)$ is known are provided in Table~\ref{tab:estres} for the two
different censoring times. Table~\ref{tab:estres2} provides the
results for the semi-parametric estimation procedure with the six
different amounts of smoothing for the $T=500$ censoring time only. The
semi-parametric algorithm is considered converged when the value of the 
log-likelihood
between consecutive iterations does not change by 
more than the tolerance $\epsilon = 0.001$.
Both tables contain the true parameters used to simulate the catalog,
the average of the estimated parameters (Est), the standard
deviation of the parameter estimates (SE), the average of the
standard error estimates obtained by inverting the observed
information matrix ($\widehat{\text{SE}}$), and the empirical coverage
probability (CP) of the 1000 approximate 95\% confidence intervals
obtained by assuming asymptotic normality of the estimators.

\begin{table}[htp!]
	\centering
	\caption{
		Estimation results for the simulated earthquake
		catalogs assuming the true form of the spatial intensity function
		$\nu(x,y)$ is known for the two censoring times
		$T=250$ and $T=500$.
		\label{tab:estres}}
	
	\vspace{0.4cm}
	
	\begin{tabular}{|ll|cc | cc | cc |cc |}
		\hline
		& & $\kappa$ & $\beta$ & $p$ & $c$ & $\sigma_1^2$ & $\sigma_2^2$ & $A$ & $\alpha$ \\ 
		& & 0.80 & 1.25 & 1.20 & 0.01 & 0.01 & 0.02 & 0.50 & 1.00 \\ 
		\hline
		\multirow{4}{*}{\rotatebox[origin=c]{90}{$T = 250$}} & Est &  0.818 & 1.254 & 1.224 & 0.0116 & 0.0107 & 0.0214 & 0.523 & 0.984  \\ 
		& SE & 0.097 & 0.203 & 0.091 & 0.0050 & 0.0020 & 0.0037 & 0.149 & 0.342 \\ 
		& $\widehat{\text{SE}}$ & 0.097 & 0.217 & 0.091 & 0.0046 & 0.0015 & 0.0030 & 0.146 & 0.358 \\ 
		& CP & 0.961 & 0.935 & 0.966 & 0.942 & 0.890 & 0.913 & 0.932 & 0.967\\ 
		\hline
		\multirow{4}{*}{\rotatebox[origin=c]{90}{$T = 500$}} & Est & 0.812 & 1.250 & 1.213 & 0.0108 & 0.0103 & 0.0209 & 0.509 & 0.994 \\ 
		& SE & 0.070 & 0.155 & 0.062 & 0.0031 & 0.0011 & 0.0025 & 0.083 & 0.240 \\
		& $\widehat{\text{SE}}$ & 0.069 & 0.161 & 0.059 & 0.0029 & 0.0010 & 0.0021 & 0.086 & 0.243 \\ 
		& CP & 0.951 & 0.935 & 0.955 & 0.945 & 0.923 & 0.901 & 0.938 & 0.961 \\ 
		\hline		
	\end{tabular}
\end{table}

Table~\ref{tab:estres} shows that when the true spatial
intensity function $\nu$ is known, the empirical biases of the
estimators are negligible compared to their respective SEs,
the estimated SEs are fairly close to the true (empirical) SEs, and
the empirical CPs of the confidence intervals are fairly close to the
nominal level of 0.95, for all parameters except $\sigma_1^2$ and
$\sigma_2^2$. The less than ideal performance of the
estimators of $\sigma_1^2$ and $\sigma_2^2$ 
might be caused by a flat log-likelihood surface in the direction of 
these parameters. However, even for such
parameters, the empirical biases and SEs decrease as expected when
$T$ increases. We next investigate the estimation results when the spatial intensity
function is also estimated using a weighted 2d-KDE. 
The estimation results summarized in Table~\ref{tab:estres2} have
been trimmed to remove unusual estimates. 
The Mahalanobis distance between each of the estimates and the true parameter 
$\theta$ was calculated, and those with the largest 5\% of distances were removed 
from the estimation results presented in the table.

\begin{table}[htp!]
	\centering
	\caption{Estimation results for the simulated earthquake catalogs with spatial intensity function estimated via a weighted 2d-KDE using the semi-parametric iterative estimation procedure with six different levels of smoothing by multiplying the default smoothing matrix by factors of $\zeta$ in the set $\set{0.5,1,1.5,2,2.5,3}$ for the censoring time $T=500$. \label{tab:estres2}}
	
	\vspace{0.4cm}
	
	\resizebox{\linewidth}{!}{
		\begin{tabular}{|l l|cc | cc | cc |cc |}
			\hline
			& & $\kappa$ & $\beta$ & $p$ & $c$ & $\sigma_1^2$ & $\sigma_2^2$ & $A$ & $\alpha$ \\ 
			& & 0.80 & 1.25 & 1.20 & 0.01 & 0.01 & 0.02 & 0.50 & 1.00 \\ 
			\hline
			\multirow{4}{*}{\rotatebox[origin=c]{90}{$\zeta = 0.5$}} &    Est & 0.835 & 0.994 & 1.428 & 0.0181 & 0.0099 & 0.0200 & 0.357 & 1.023\\ 
			& SE &  0.057 & 0.091 & 0.081 & 0.0048 & 0.0011 & 0.0022 & 0.038 & 0.248 \\
			& $\widehat{\text{SE}}$ & 0.056 & 0.089 & 0.090 & 0.0049 & 0.0010 & 0.0020 & 0.038 & 0.249  \\ 
			& CP & 0.931 & 0.237 & 0.155 & 0.7733 & 0.9131 & 0.9206 & 0.070 & 0.958 \\ 
			\hline
			\multirow{4}{*}{\rotatebox[origin=c]{90}{$\zeta = 1$}} &  Est &  0.835 & 1.053 & 1.352 & 0.0155 & 0.0100 & 0.0202 & 0.391 & 1.014  \\ 
			& SE & 0.061 & 0.105 & 0.074 & 0.0041 & 0.0011 & 0.0022 & 0.045 & 0.245 \\ 
			& $\widehat{\text{SE}}$ &  0.060 & 0.106 & 0.080 & 0.0043 & 0.0010 & 0.0020 & 0.046 & 0.246 \\ 
			& CP & 0.935 & 0.503 & 0.549 & 0.8967 & 0.9199 & 0.9283 & 0.352 & 0.960 \\ 
			\hline
			\multirow{4}{*}{\rotatebox[origin=c]{90}{$\zeta = 1.5$}} &  Est &  0.834 & 1.098 & 1.310 & 0.0142 & 0.0100 & 0.0202 & 0.418 & 1.006 \\ 
			& SE & 0.064 & 0.117 & 0.071 & 0.0038 & 0.0011 & 0.0022 & 0.052 & 0.244 \\ 
			& $\widehat{\text{SE}}$ &   0.063 & 0.120 & 0.075 & 0.0039 & 0.0010 & 0.0020 & 0.054 & 0.245  \\ 
			& CP & 0.935 & 0.681 & 0.797 & 0.9494 & 0.9220 & 0.9231 & 0.579 & 0.962 \\ 
			\hline
			\multirow{4}{*}{\rotatebox[origin=c]{90}{$\zeta = 2$}} & Est. &  0.832 & 1.148 & 1.275 & 0.0131 & 0.0100 & 0.0202 & 0.448 & 0.999 \\ 
			& SE&   0.067 & 0.133 & 0.068 & 0.0036 & 0.0011 & 0.0022 & 0.061 & 0.242 \\ 
			& $\widehat{\text{SE}}$ & 0.066 & 0.137 & 0.071 & 0.0036 & 0.0010 & 0.0020 & 0.064 & 0.244 \\ 
			& CP & .940 & 0.807 & 0.903 & 0.9673 & 0.9178 & 0.9252 & 0.774 & 0.962 \\ 
			\hline
			\multirow{4}{*}{\rotatebox[origin=c]{90}{$\zeta = 2.5$}} & Est &  0.829 & 1.213 & 1.240 & 0.0120 & 0.0100 & 0.0202 & 0.488 & 0.990\\ 
			& SE & 0.071 & 0.156 & 0.065 & 0.0033 & 0.0011 & 0.0022 & 0.078 & 0.242\\ 
			& $\widehat{\text{SE}}$ & 0.070 & 0.159 & 0.067 & 0.0033 & 0.0010 & 0.0020 & 0.080 & 0.242  \\ 
			& CP & 0.941 & 0.887 & 0.964 & 0.9673 & 0.9199 & 0.9262 & 0.910 & 0.956 \\ 
			\hline  
			\multirow{4}{*}{\rotatebox[origin=c]{90}{$\zeta = 3$}} & Est &   0.824 & 1.307 & 1.201 & 0.0108 & 0.0100 & 0.0202 & 0.556 & 0.981\\ 
			& SE & 0.076 & 0.193 & 0.064 & 0.0031 & 0.0010 & 0.0021 & 0.126 & 0.243 \\ 
			& $\widehat{\text{SE}}$ &   0.075 & 0.192 & 0.061 & 0.0030 & 0.0010 & 0.0020 & 0.113 & 0.241  \\ 
			& CP &   0.946 & 0.948 & 0.922 & 0.9336 & 0.9283 & 0.9283 & 0.979 & 0.953  \\ 
			\hline
	\end{tabular}}
\end{table}

It should be expected that different amounts of smoothing used in
estimating the spatial intensity function introduce different biases
in the estimation of the \retas\ model parameters.
From Table~\ref{tab:estres2} we note a clear positive correlation between
the amount of smoothing and the bias of the estimator of $A$ and
negative correlation between smoothness and the bias of the estimator
of $p$. This is to be expected since less smoothing in the
background intensity estimator $\hat\nu(x,y)$ leads to a more
flexible background spatial intensity function, which might explain
away some of the short-range spatial variation among the 
induced/excited earthquakes and therefore less earthquakes will be
attributed to excitation effect, and the ones that are attributed to
excitation effect will also appear to be more tightly clustered in
space and time. This in turn will lead to smaller estimates values of 
$A$, and larger estimates of $p$, since a smaller $A$ implies a
fewer number of aftershocks due to an earthquake and a larger $p$
indicates a faster decaying excitation effect. Table~\ref{tab:estres2}
also reveals a clear positive correction between the amount of
smoothing and the bias in the estimator of $\beta$, which is to be
expected since less smoothing causes more earthquakes to be classified
as main-shocks and therefore the average waiting time between main
shocks will appear smaller, which leads to smaller values of the scale
parameter $\beta$ for the main-shock waiting time distribution. The
relationships between smoothness of $\hat\nu(x,y)$ and the biases of
other parameters can be explained similarly, although they are less
pronounced.

The objective determination of the appropriate amount of smoothing will 
be achieved using the AICc criteria as discussed in Subsection~\ref{sec:smoothing}. 
For this purpose, 
we first calculate the effective number of parameters used in the estimation 
of the spatial intensity function. As the amount of smoothing increases, 
the effective number of parameters decreases. For instance, the average 
number of parameters for the estimated spatial intensity functions 
obtained from the simulated catalogs are 
85.95, 47.32, 32.37, 24.37, 19.39 and 16.01 for $\zeta = 0.5,1,1.5,2,2.5$ 
and $3$ respectively. 
The mean AICc value for the six different amounts of smoothing
are -0.62, -28.44, -32.77, -29.92, -23.65, and -15.43 for
$\zeta =0.5,1,1.5,2,2.5$ and $3$ respectively. This indicates that the 
default smoothing matrix of \cite{Wand1994} multiplied by factors between 
$\zeta = 1$ and $\zeta = 2$ tend to perform best in terms of the AICc 
information criteria. Indeed, the multiplication factors
$\zeta = 1, 1.5$ and $2$ were selected 6.88\%, 85.24\% and 7.89\% 
of the time, respectively.  

The parameter estimates based on the simulated catalogs with the amount of smoothing 
selected by the AICc are presented in Table~\ref{tab:estres3}, where again the estimation 
results were trimmed using the same strategy used in Table~\ref{tab:estres2}.
The table shows that the biases of the estimators, the biases of
the standard error estimators, and the coverage probabilities of the
confidence intervals are all comparable with those
obtained by using the known form of the spatial intensity function during the estimation
process. The estimation results could be improved further by fitting \retas\ models
based on a finner grid of $\zeta$ values, rather than increments of 
$0.5$. Therefore, we suggest that the AICc based model selection procedure 
provides a reasonable strategy for selecting the appropriate amount of smoothing.

\begin{table}[htp!]
	\centering
	\caption{Estimation results for the simulated earthquake catalogs with the amount of smoothing determined by minimizing the AICc value to estimate the spatial intensity function based on a weighted 2d-KDE using the semi-parametric iterative estimation procedure. \label{tab:estres3}}
	
	\vspace{0.4cm}
	
	\begin{tabular}{|l | cccccccc|}
		\hline
		& $\kappa$ & $\beta$ & $p$ & $c$ & $\sigma_1^2$ & $\sigma_2^2$ & $A$ & $\alpha$ \\ 
		\hline
		True & 0.800 & 1.250 & 1.200 & 0.0100 & 0.0100 & 0.0200 & 0.500 & 1.000 \\ 
		Est. & 0.823 & 1.305 & 1.210 & 0.0111 & 0.0100 & 0.0202 & 0.553 & 0.982 \\  
		SE &  0.077 & 0.214 & 0.079 & 0.0034 & 0.0011 & 0.0021 & 0.140 & 0.242 \\ 
		$\widehat{\text{SE}}$ &  0.075 & 0.192 & 0.062 & 0.0030 & 0.0010 & 0.0020 & 0.112 & 0.241 \\ 
		CP & 0.946 & 0.917 & 0.881 & 0.931 & 0.921 & 0.927 & 0.927 & 0.956 \\ 
		\hline
	\end{tabular}
\end{table}

\subsection{Declustering}
\label{subsec:declustering}
Declustering is important not only for estimation of the \retas\ model but also for 
accurate inferring of the branching structure of an earthquake catalog.
To assess the accuracy of the declustering algorithm, when simulating the earthquake
catalogs, we retain the main and aftershock labels, which are
typically not observed in practice.
We then apply the semi-parametric estimation procedure and then from the resultant 
estimates apply the declustering algorithm discussed in
Section~\ref{subsec:declustering} to $N = 1000$ simulated catalogs
according to the same simulation model as before but with seven
different parameter specifications given by:
$\kappa \in \set{0.2,0.3,0.5,1,2,3,5}$, $\beta = 1/\kappa$, $p=1.2$,
$c=0.01$, $\sigma_1^2 = 0.01$, $\sigma_2^2 = 0.02$, $A=0.5$ and
$\alpha = 1$ with censoring time $T=250$ and
$\mathcal{S} = \mathbb{R}^2$. The spatial intensity function $\nu(x,y)$ was 
again bivariate normal with independent marginals with variances
0.05 and 0.10 in the $x$-direction and $y$-direction respectively.
For each earthquake $i$ in a simulated catalog, we apply the
declustering algorithm to calculate estimates of the main-shock
probability $\omega_i$ and the triggering (parent) probabilities 
$\pi_{ij}$ of $i$ being induced by a previous earthquake 
$j\in\set{1,\dotsc,i-1}$. We compare three different situations:
\begin{itemize}
	\item Using the \retas\ model and the declustering algorithm based on the smoothed branching structure probabilities.
	\item Using the \retas\ model and the declustering algorithm based on the filtered branching structure probabilities. 
	\item Using the \etas\ model with the declustering algorithm based on the filtered (and equivalently smoothed) branching structure probabilities. 
\end{itemize}

\begin{figure}[h!]
	\centering
	\parbox{6cm}{
		\includegraphics[width=\linewidth]{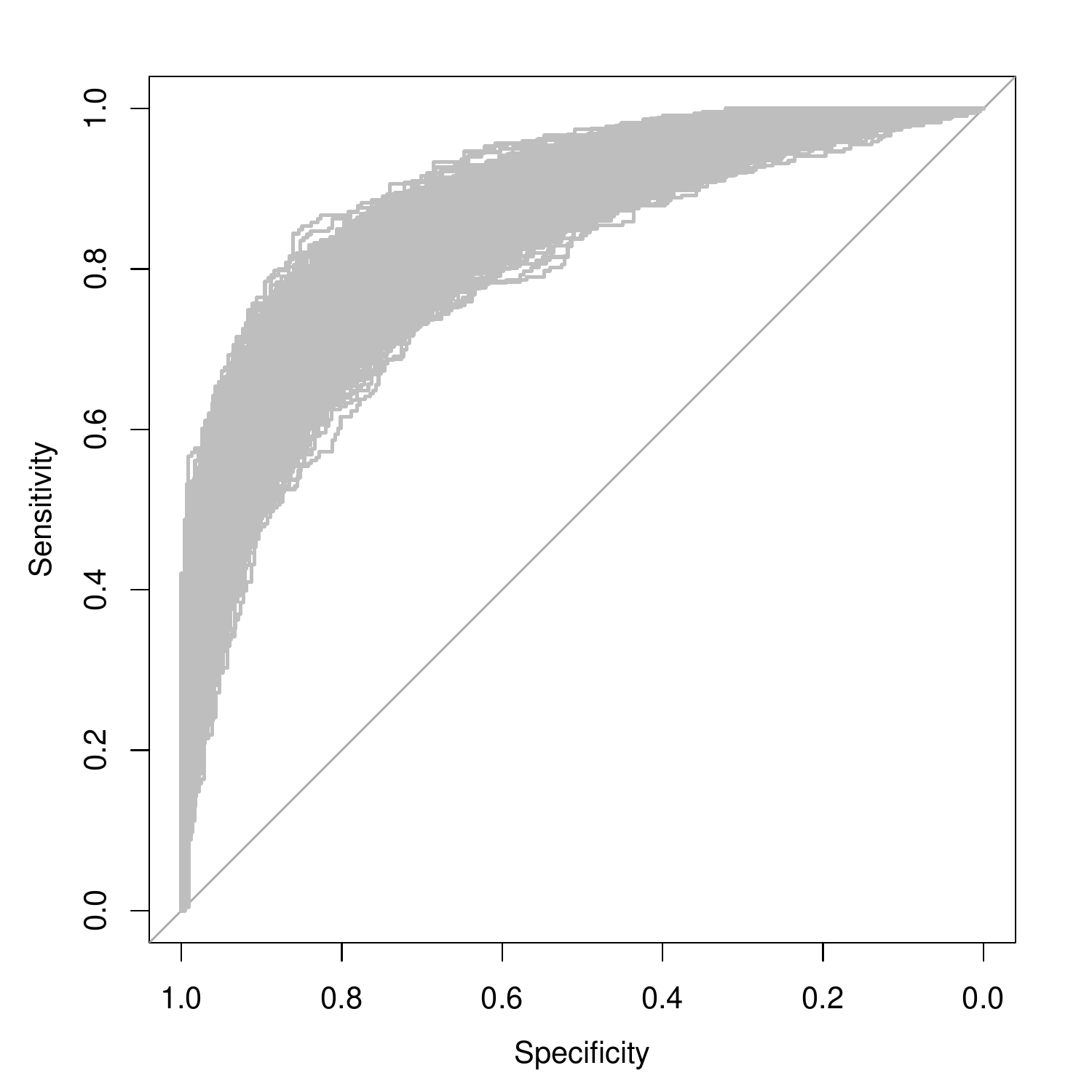}
	}
	\qquad
	\begin{minipage}{6cm}
		\includegraphics[width=\linewidth]{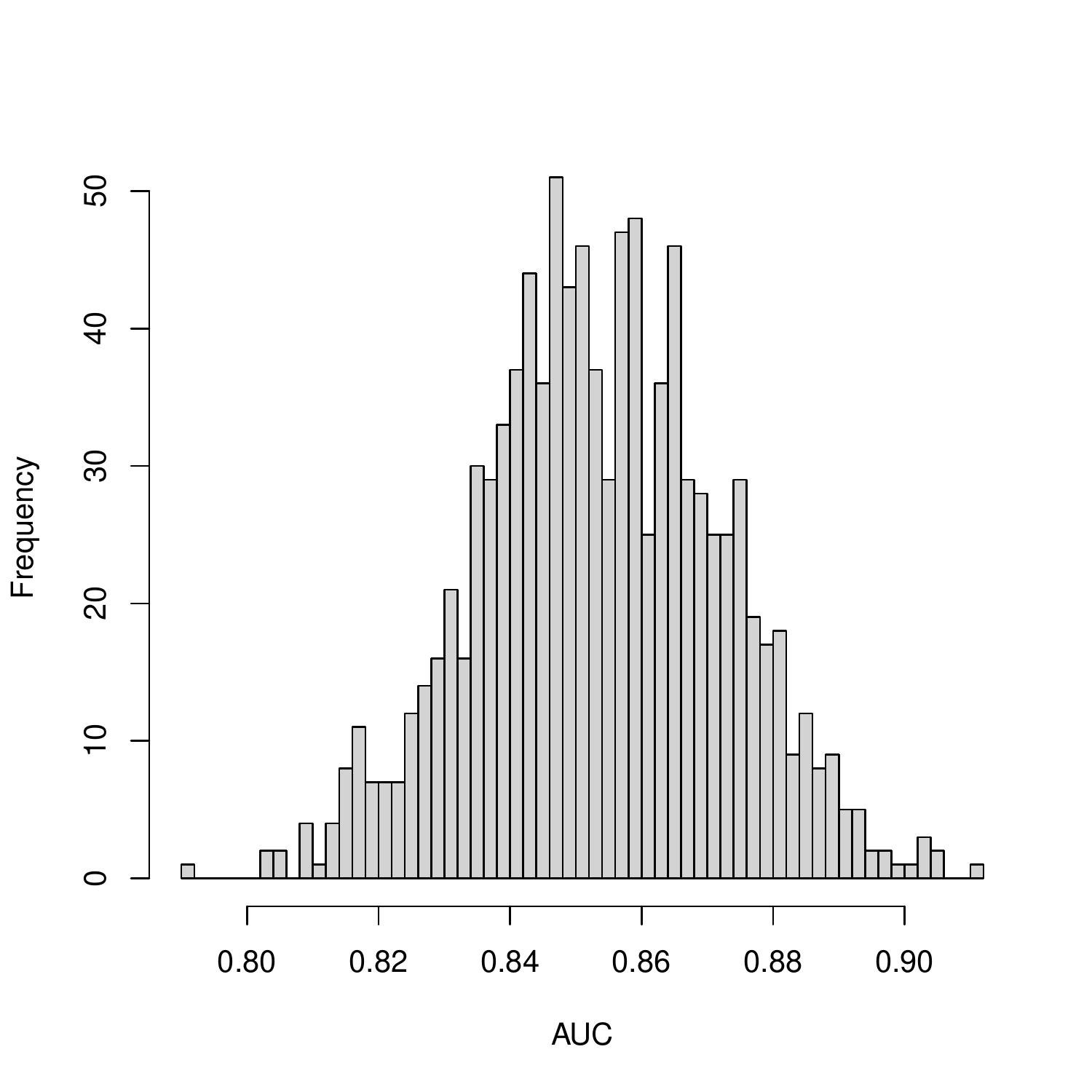}
	\end{minipage}
	\caption{\label{fig:ROC-MS} ROC curves (left plot) for the classification of
		main-shocks using the main-shock probabilities $\omega_i$ derived
		from the \retas\ declustering algorithm applied to the simulated
		earthquake catalogs for the case where $\kappa = 0.5$ and
		$\beta = 2$. Histogram (right plot) of the AUC values for the ROC curves in the left plot.}
\end{figure}

Classification of an earthquake as either a main-shock or an aftershock is
a binary classification problem, and therefore, the receiver
operating characteristic (ROC) curve can be used to compare the
estimated main-shock probabilities $\omega_i$
with the true main-shock labels. For each of the three situations described above, 
we calculate the ROC curves 
for all 1000 simulated catalogs and seven parameter specifications, 
and only show those for the \retas\ model with $\kappa = 0.5$ and $\beta = 2$ 
in Figure~\ref{fig:ROC-MS} (left panel) using the declustering algorithm with the smoothed probabilities. 
The areas under the ROC curves (AUC) were calculated as well. 
The right panel of Figure~\ref{fig:ROC-MS} presents
a histogram of the AUC values, which shows the consistently high 
AUC values and therefore the good performance of the declustering 
algorithm for main or aftershock classification.


The declustering algorithm provides accurate identification of
main-shocks (hence also aftershocks) since consistently 
high AUC values are observed for the range of parameterizations 
\begin{table}[h!]
	\caption{\label{tab:AUC-values} Summary statistics for the AUC
		values for 1000 simulated earthquake catalogs with shape parameter
		$\kappa$ in the set $\set{0.2, 0.3,0.5,1,2,3,5}$.  The \retas\ declustering
		algorithm was used to estimate the main-shock probabilities
		$\omega_i$ for the \retas\ and then the \etas\ declustering algorithm was 
		used for both the \retas\ and \etas\ models. }
	
	\vspace{0.4cm}
	\centering
	\begin{tabular}{|l l | cccccc|}
		\hline
		& & Min & 1st Qu. & Median & Mean & 3rd. Qu & Max. \\
		\hline
		\multirow{3}{*}{$\kappa = 0.2$} & RETAS & 0.8111 & 0.8742 & 0.8860 & 0.8848 & 0.8963 & 0.9263  \\
		& Filtered & 0.7294 & 0.8004 & 0.8140 & 0.8137 & 0.8288 & 0.8783 \\ 
		& ETAS & 0.5059 & 0.5814 & 0.6049 & 0.6067 & 0.6290 & 0.7306\\ 
		\hline
		\multirow{3}{*}{$\kappa = 0.3$} & RETAS & 0.8229 & 0.8623 & 0.8733 & 0.8727 & 0.8841 &  0.9137  \\ 
		& Filtered & 0.7598 & 0.8258 & 0.8395 & 0.8385 & 0.8511 & 0.8952 \\ 
		& ETAS & 0.6053 & 0.7392 & 0.7589 & 0.7571 & 0.7767 & 0.8366 \\ 
		\hline
		\multirow{3}{*}{$\kappa = 0.5$} & RETAS & 0.7904 & 0.8413 & 0.8527 & 0.8536 & 0.8660 &  0.9108 \\ 
		& Filtered & 0.7864 & 0.8309 & 0.8428 & 0.8430 & 0.8555 & 0.9058  \\ 
		& ETAS & 0.7576 & 0.8110 & 0.8240 & 0.8239 & 0.8372 & 0.8828 \\ 
		\hline
		\multirow{3}{*}{$\kappa = 1$} & RETAS & 0.7922 & 0.8531 & 0.8657 & 0.8647 & 0.8766 & 0.9151  \\  
		& Filtered & 0.7915 & 0.8532 & 0.8659 & 0.8649 & 0.8767 & 0.9153  \\ 
		& ETAS &0.7903 & 0.8537 & 0.8662 & 0.8652 & 0.8771 & 0.9156 \\ 
		\hline
		\multirow{3}{*}{$\kappa = 2$} & RETAS & 0.8370 & 0.8859 & 0.8956 & 0.8951 & 0.9052 & 0.9476 \\ 
		& Filtered &  0.8295 & 0.8788 & 0.8885 & 0.8883 & 0.8984 & 0.9457\\
		& ETAS & 0.8213 & 0.8722 & 0.8829 & 0.8826 & 0.8936 & 0.9416 \\ 
		\hline
		\multirow{3}{*}{$\kappa = 3$} & RETAS & 0.8666 & 0.9046 & 0.9141 & 0.9138 & 0.9236 & 0.9483  \\ 
		& Filtered & 0.8523 & 0.8911 & 0.9015 & 0.9009 & 0.9116 & 0.9432 \\ 
		& ETAS &  0.8370 & 0.8786 & 0.8891 & 0.8889 & 0.8998 & 0.9362 \\ 
		\hline
		\multirow{3}{*}{$\kappa = 5$} & RETAS &0.8809 & 0.9258 & 0.9339 & 0.9334 & 0.9419 & 0.9677 \\ 
		& Filtered & 0.8616 & 0.9026 & 0.9128 & 0.9117 & 0.9213 & 0.9523  \\ 
		& ETAS & 0.8402 &  0.8828  &0.8935 & 0.8927 & 0.9028 & 0.9354 \\ 
		\hline
	\end{tabular}
\end{table}
as presented in Table~\ref{tab:AUC-values}. For comparison, we also 
present the results of the declustering algorithm based on the \etas\ model. 
When $\kappa$ departs from one, the \retas\ model consistently performs
better at identifying main-shocks and has a more pronounced improvement when
$\kappa$ departs further from one.  For instance, when
$\kappa = 0.2$, the mean AUC for the \retas\ model is $0.88$ while for
the \etas\ model it was significantly smaller at only $0.61$. When
$\kappa = 1$ the two models perform similarly as one would expect
since the two models are very similar in this case (and would be identical 
if we imposed the restriction that $\hat\kappa = 1$). The table also 
contains the results if the \etas\ declustering algorithm discussed in 
Section~\ref{sec:filtered} was applied for \retas\ model declustering. 
The filtered probabilities have consistently poorer performance than 
the declustering algorithm based on the smoothed probabilities except 
when $\kappa = 1$. However, the table suggest that selecting the correct 
model is more important than the declustering method used.

We also need to assess whether the declustering algorithm can recover the complete
branching structure that includes the parent for each aftershock. 
For this purpose, we compare the main and aftershock probabilities 
derived from the declustering algorithm to the true 
simulated branching structure consisting of main-shock and parent labels.
The most probable label classification for each earthquake $i=2,\dotsc,n$ based
\begin{table}[h!]
	\caption{\label{tab:percent-correct} Summary statistics for
		the proportion of the entire branching structure inferred
		correctly (based on the most probable classification) for 1000 simulated earthquake catalogs with shape
		parameter $\kappa$ in the set $\set{0.2, 0.3,0.5,1,2,3,5}$.  
		The \retas\ declustering
		algorithm was used to estimate the main-shock
		and parent probabilities
		$\set{\omega_i,\pi_{i,j},j=1,\dotsc, i-1}$ for the \retas\ and then the \etas\ declustering algorithm was 
		used for both the \retas\ and \etas\ models.
	}
	
	\vspace{0.4cm}
	
	\centering
	\begin{tabular}{|l l | cccccc|}
		\hline
		& & Min. & 1st Qu. & Median & Mean & 3rd Qu. & Max. \\ 
		\hline
		\multirow{3}{*}{$\kappa = 0.2$} & RETAS & 0.5793 & 0.6320 & 0.6491 & 0.6494 & 0.6653 &  0.7408 \\ 
		& Filtered & 0.5356 & 0.6031 & 0.6213 & 0.6215 & 0.6385 & 0.7242  \\ 
		& ETAS & 0.4097 & 0.4654 & 0.4812 & 0.4827 & 0.5000 & 0.5787  \\ 
		\hline
		\multirow{3}{*}{$\kappa = 0.3$} & RETAS & 0.6139 & 0.6612 & 0.6786 & 0.6785 & 0.6948 &  0.7566  \\ 
		& Filtered & 0.5955 & 0.6485 & 0.6653 & 0.6656 & 0.6816 & 0.7500  \\ 
		& ETAS & 0.4887 & 0.5782 & 0.5970 & 0.5972 & 0.6153 & 0.6901 \\ 
		\hline
		\multirow{3}{*}{$\kappa = 0.5$} & RETAS & 0.5960 & 0.6694 & 0.6864 & 0.6870 & 0.7036 &  0.7833 \\ 
		& Filtered & 0.5900 & 0.6661 & 0.6831 & 0.6839 & 0.6999 & 0.7757 \\ 
		& ETAS & 0.5820 & 0.6515 & 0.6692 & 0.6692 & 0.6864 & 0.7573  \\ 
		\hline
		\multirow{3}{*}{$\kappa = 1$} & RETAS & 0.6252 & 0.6936 & 0.7123 & 0.7111 & 0.7273 & 0.8018 \\
		& Filtered & 0.6269 & 0.6938 & 0.7119 & 0.7111 & 0.7273 & 0.7995 \\ 
		& ETAS & 0.6269 & 0.6946 & 0.7121 & 0.7115 & 0.7275 & 0.7995  \\ 
		\hline
		\multirow{3}{*}{$\kappa = 2$} & RETAS & 0.6551 & 0.7137 & 0.7322 & 0.7317 & 0.7490 & 0.8124 \\ 
		& Filtered & 0.6475 & 0.7124 & 0.7303&  0.7298&  0.7471&  0.8101 \\
		& ETAS &0.6460 & 0.7066 & 0.7238 & 0.7238 & 0.7407 & 0.7963  \\ 
		\hline
		\multirow{3}{*}{$\kappa = 3$} & RETAS & 0.6719 & 0.7260 & 0.7443 & 0.7442 & 0.7614 & 0.8434 \\ 
		& Filtered & 0.6744 & 0.7241 & 0.7402 & 0.7404 & 0.7569 & 0.8434 \\ 
		& ETAS & 0.6609 & 0.7132 & 0.7305 & 0.7298 & 0.7459 & 0.8361\\ 
		\hline
		\multirow{3}{*}{$\kappa = 5$} & RETAS & 0.6797 & 0.7381 & 0.7544 & 0.7547 & 0.7709 & 0.8390  \\ 
		& Filtered & 0.6705 & 0.7284 & 0.7453 & 0.7456 & 0.7621 & 0.8277\\ 
		& ETAS & 0.6606 & 0.7142 & 0.7305 & 0.7304 & 0.7460 & 0.8186 \\ 
		\hline
	\end{tabular}
\end{table}
on $\max \set{\omega_i, \pi_{ij}, j=1,\dotsc, i-1}$ 
is computed and compared with the true label from the
simulations
. Table~\ref{tab:percent-correct} presents the proportion of
the branching structure correctly
inferred where a match is defined as the most probable index
coinciding with the true parent or main-shock label. 
The \retas\ delcustering algorithm based on the smooth probabilities 
correctly infers the majority of the
branching structure with a mean proportion around $0.70$ for all parameter
specifications under consideration. In some cases, the declustering
algorithm can infer as much as $84\%$ of the branching structure
correctly. By comparing with the \etas\ model, it always has a superior 
performance and the largest improvement is seen when $\kappa$ is smallest 
($\kappa = 0.2$) in which there is a significant reduction of $0.17$ in the mean 
correct proportion inferred. Similar to Table~\ref{tab:AUC-values}, the filtered 
declustering algorithm has worse performance for all parameter specifications 
(when $\kappa$ departs from one) but its performance is still consistently better
than the \etas\ model results again suggesting that getting the model correct 
is more important.

\section{Application to a New Zealand earthquake catalog}
\label{sec:applications}

This section investigates a New Zealand (NZ) earthquake catalog from
1-Jan-1980 to 29-Feb-2020. The catalog was obtained from the GeoNet 
Quake Search Database and reports the hypocentral coordinates 
(longitude, latitude and time) and magnitude for 1173 
earthquakes with threshold magnitude $m_0 = 5$.  The coordinates $164^\circ-182^\circ$~E and $48^\circ - 35 ^\circ$~S define the 
rectangular region $\mathcal S$, which was previously studied 
in~\cite{Harte2013, Harte2014} and  \cite{Stindl2021}. The region covers 
all of New Zealand and hence different tectonic environments exists 
within the catalog.

\begin{figure}[t!]
	\centering
	\includegraphics[width=1\textwidth]{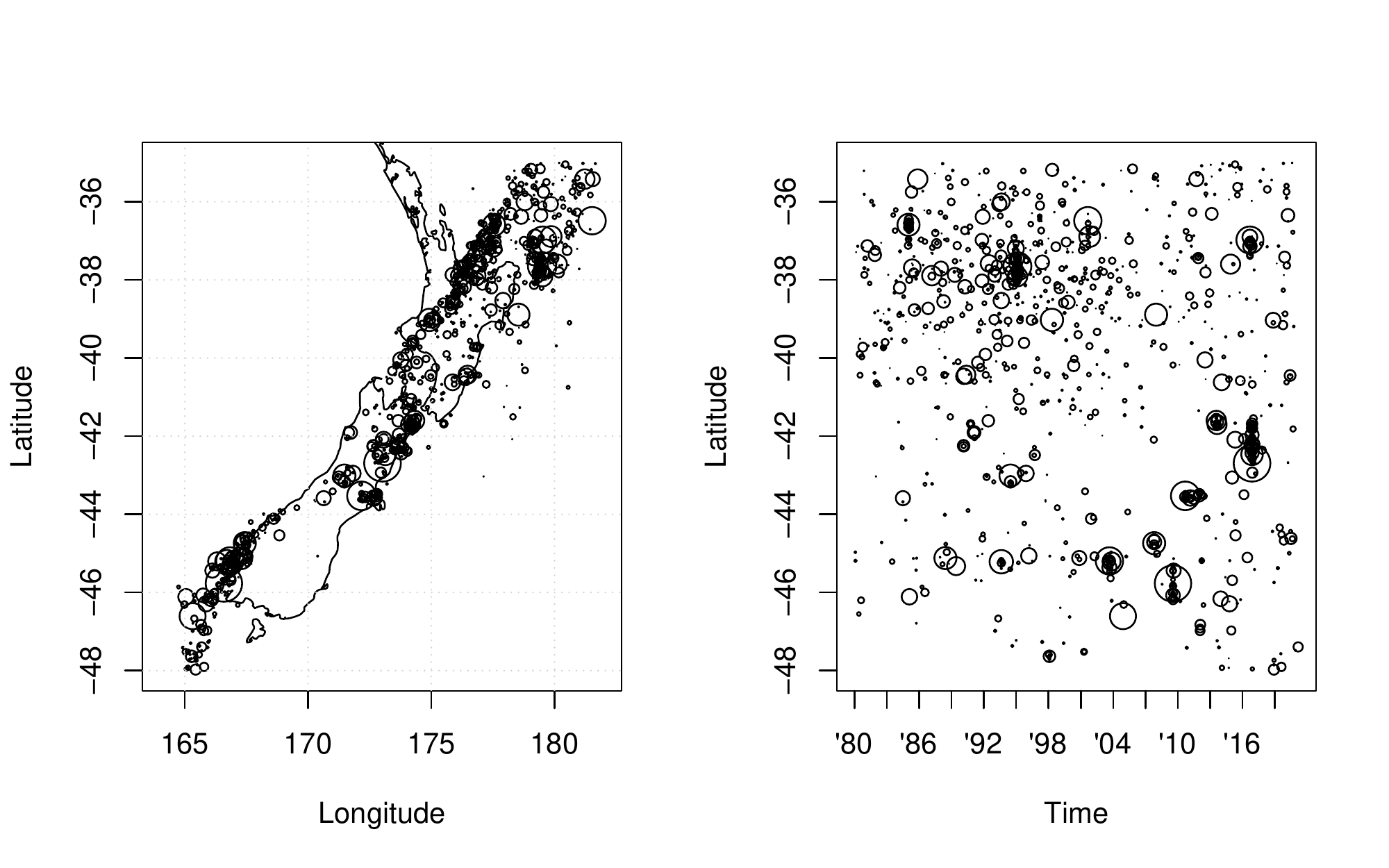}
	\caption{Epicentral locations (left plot) and latitude-time (right plot) plots for the NZ catalog 
		where the relative size of the magnitude is indicated by the size of the circle. \label{fig:NZplot} }
\end{figure}

Figure~\ref{fig:NZplot} displays the epicentral locations (left plot)
and longitude-time occurrences (right plot) of earthquakes in the NZ 
catalog where the relative size of the magnitude is represented 
by the size of the circle. The figure shows strong evidence of 
heavy temporal and spatial clustering of the earthquakes. Seismic 
activity typically occurs where the Pacific Plate is subducting the 
Australian Plate in both the North Island and in the northern part 
of the South Island, the Australian Plate is subducting the Pacific 
Plate in the southwest of the South Island and the Alpine Fault 
which is located in the central part of the South Island 
\citep{Harte2013}. The seismic activity  
is greatest closest to the tectonic boundaries and diminishes as one moves 
from the boundary to the west and east into the Tasman Sea and 
Pacific Ocean, respectively.

The specific form of the \retas\ model for this catalog is the same as
described in Section~\ref{sec:RETAS-Model}. The semi-parametric
iterative estimation algorithm is used to estimate the spatial
intensity function $\nu(x,y)$ and the model parameters
$\theta = (\kappa, \beta, p, c, \sigma_1^2, \sigma_2^2, A, \alpha)$, with
the default smoothing parameter matrix multiplied by the factors
$\zeta \in \set{0.1, 0.25, 0.5,1,2,3,4}$.  The spatial intensity
function is initialized using a 2d-KDE with equal weights assigned to
all earthquakes in the catalog.  The optimization routine to obtain 
the \retas\ model parameters was
initialized by fitting telescopically simpler models nested in the
\retas\ model. 
The parameter estimates, standard errors, productivity (Prod.),
percentage of expected main-shocks (Pct), log-likelihood $\ell(\hat\theta)$, 
effective number of parameters DoF for $\nu(x,y)$ and the AICc are provided in
Table~\ref{tab:NZ-res} for each adjustment factor. The magnitude density 
term in the likelihood function is separable, and its value is 
$-1410.97$ with $\hat \gamma = 0.336$ and is the same in all cases 
and is not included in the tables calculation of $\ell(\hat \theta)$.

\begin{landscape}
\begin{table}[htp!]
	\centering
	\caption{Estimation results for the
		semi-parametric estimation procedure with six different
		amounts of smoothing by multiplying 
		the default smoothing matrix $h$ by factors of $\zeta$ 
		in the set $\set{0.1,25,0.5,1,2,3,4}$ for the NZ catalog.  
		\label{tab:NZ-res}}
	
	\vspace{0.4cm} 
	
	\resizebox{1.5\textwidth}{!}{
		\begin{tabular}{|c | cccccccc|cc|cccc|}
			\hline
			$\zeta$ & $\kappa$ & $\beta$ & $p$ & $c$ & $\sigma_1^2$ & $\sigma_2^2$ & $A$ & $\alpha$ & Prod. & Pct &$\ell(\hat \theta)$ &DoF & AICc & \\ 
			\hline
			\multirow{2}{*}{0.1} & 0.853 & 25.61 &  1.165 &  0.00601 &  0.0168 &  0.00898 &  0.213 & 1.538 &  \multirow{2}{*}{0.442} & \multirow{2}{*}{58.63} & \multirow{2}{*}{-4943.33} & \multirow{2}{*}{362.62} & \multirow{2}{*}{10845.20}  & \\
			&  (0.0426) & (1.693) & (0.0219) & (0.00135) & (0.00165) & (0.00102) & (0.0196) & (0.0780) &&&&&&\\
			\hline
			\multirow{2}{*}{0.25} & 0.850 &  26.50 &  1.136 &  0.00508 &  0.0170 &  0.00900 & 0.242 & 1.498 &  \multirow{2}{*}{0.488} & \multirow{2}{*}{56.84} &  \multirow{2}{*}{ -5095.81} & \multirow{2}{*}{214.36} & \multirow{2}{*}{10721.73} &  \\
			& (0.0446) & (1.908) & (0.0191) & (0.000993) & (0.00168) & (0.00105) & 
			(0.0294) &  (0.0773) &&&&&&\\
			\hline
			\multirow{2}{*}{\textbf{0.5}} & \textbf{0.848} &  \textbf{27.25} &  \textbf{1.115} &  \textbf{0.00450} &  \textbf{0.0172} & \textbf{0.00911} &  \textbf{0.264} &  \textbf{1.506} & \multirow{2}{*}{\textbf{0.534}} & \multirow{2}{*}{\textbf{55.41}} & \multirow{2}{*}{\textbf{-5192.18}}  & \multirow{2}{*}{\textbf{132.87}} & \multirow{2}{*}{\textbf{10709.65}} & \\
			& \textbf{(0.0446)} & \textbf{(1.908)} & \textbf{(0.0191)} & \textbf{(0.000993)} & \textbf{(0.00168)} & \textbf{(0.00105)} & 
			\textbf{(0.0294)} &  \textbf{(0.0773)} &&&&&&\\
			\hline		
			\multirow{2}{*}{1} & 0.844 &  28.26 &  1.091 &  0.00383 &  0.0174 & 0.009216 &  0.308 &  1.488 & \multirow{2}{*}{0.616} & \multirow{2}{*}{53.66} &  \multirow{2}{*}{-5274.37}  &  \multirow{2}{*}{79.44}& \multirow{2}{*}{10745.47} &  \\
			& (0.0458) & (2.054) & (0.0186) & (0.000833) &  (0.00170) &
			(0.00107) & (0.0432) & (0.0772) &&&&&&\\
			\hline
			\multirow{2}{*}{2} & 0.839 & 29.58 & 1.0688 &  0.00328 &  0.0174 & 0.009315 &  0.384 &  1.443 & \multirow{2}{*}{0.738} & \multirow{2}{*}{51.63} &  \multirow{2}{*}{-5346.27} & \multirow{2}{*}{45.65} & \multirow{2}{*}{10812.23} &  \\
			& (0.0473) & (2.251) & (0.0173) & (0.000678) & (0.00172) &
			(0.00108) & (0.0700) & (0.0783)&&&&&&\\
			\hline
			\multirow{2}{*}{3} &0.836 & 30.63 & 1.055 & 0.0030 & 0.0175 & 0.0094 & 0.461 & 1.411& \multirow{2}{*}{0.877} & \multirow{2}{*}{50.00} & \multirow{2}{*}{-5386.93} & \multirow{2}{*}{32.42} & \multirow{2}{*}{10863.96} & \\
			& (0.0486) & (2.409) & (0.0158) & (0.000587) & (0.00172) & (0.00110) & (0.100) & (0.0788)&&&&&&\\
			\hline
			\multirow{2}{*}{4} & 0.831 & 31.70 & 1.043 &  0.00273 &  0.0177 & 0.00958 &  0.563 &  1.417 & \multirow{2}{*}{1.075} & \multirow{2}{*}{48.62} & \multirow{2}{*}{-5416.68}  & \multirow{2}{*}{25.44} & \multirow{2}{*}{10907.63} & \\
			& (0.0495) & (2.563) & (0.0147) & (0.000498) & (0.00173) &
			(0.00112) & (0.156) & (0.0775)&&&&&&\\
			\hline
	\end{tabular}}
\end{table}
\end{landscape}

Table~\ref{tab:NZ-res} exhibits similar patterns in parameter estimates to
those observed in the simulation experiments as the amount of smoothing
varies. 
The inference on the catalog changes substantially as $\zeta$
vaies. For example, the expected number of directly induced
aftershocks, aka the productivity measure (Prod. in
Table~\ref{tab:NZ-res}),
$ \mathbb{E} \big[Ae^{\alpha(m_i - m_0)}\big] = A\gamma
/(\gamma-\alpha)$ changes from the subcritical value of $0.442$ when
$\zeta=0.1$ to the supercritical $1.075$ when $\zeta=4$. 
Moreover, the percentage of main-shocks (Pct in Table~\ref{tab:NZ-res}) 
also changed considerably from 58.63\% when $\zeta = 0.1$ 
to 48.62\% when $\zeta = 4$.
The minimal AICc value is achieved when $\zeta = 0.5$.  We also
investigated more extreme amounts of smoothing in both directions, and
found that the AICc continues to increase. Therefore, we choose
$\zeta=0.5$ as the optimal amount of smoothing and proceed with the
analysis accordingly. The corresponding smoothing matrix 
$h = \bigl( \begin{smallmatrix}0.176 & 0.135\\ 0.135 & 0.118\end{smallmatrix}\bigr)$
is more or less in line with the amount of
smoothing used in the work of \cite{Harte2013} selected by more ad hoc
means.

\begin{figure}[h!]
	\centering
	\includegraphics[width=0.70\textwidth]{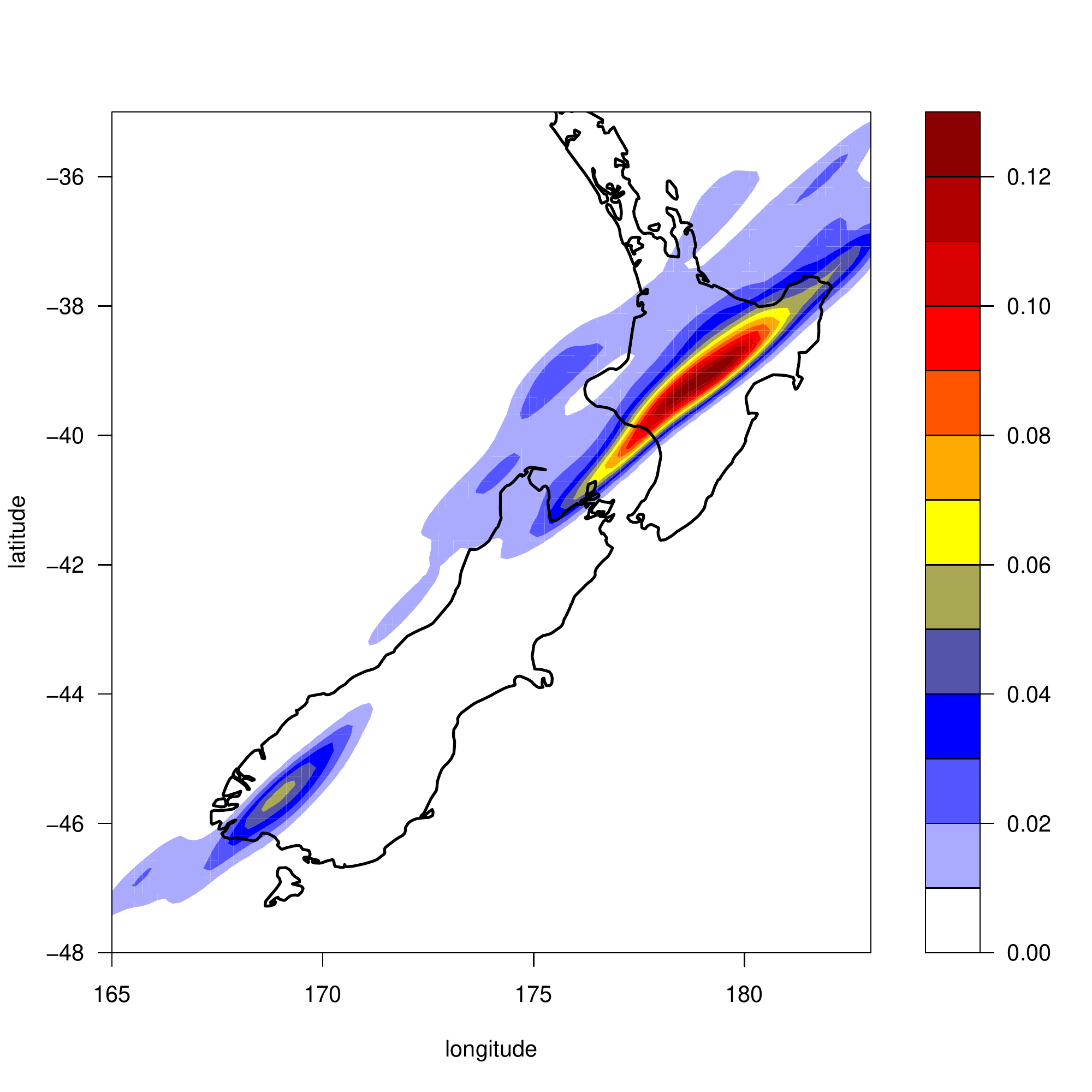}
	\caption{Estimate of the spatial intensity function $\hat \nu(x,y)$ 
		from the final iteration of the semi-parametric estimation 
		procedure with optimal amount of smoothing $\zeta = 0.5$ 
		using a weighted 2d-KDE for the NZ catalog. 
		\label{fig:NUplot} }
\end{figure}

The estimate of the main-shock spatial intensity function
$\hat{\nu}(x,y)$ is presented in Figure~\ref{fig:NUplot}. The majority
of seismic background activity is occurring along the fault lines that
go through NZ. There are two distinct seismically active
regions that have significantly high incidence rate of
main-shocks. Since we are dealing with a large space-time window, the
earthquake process will generally be more heterogeneous than smaller
space-time windows. Therefore, the rather flexible estimate for the
spatial intensity function is necessary to capture the finer features
of the fault lines. 

The estimated model parameters and standard error estimates (in brackets)
are as follows:
$\hat \kappa = 0.848 \,\, (0.0446)$,
$\hat \beta = 27.24 \,\, ( 1.908)$, $\hat p = 1.115 \,\, (0.0191)$,
$\hat c = 0.00450 \,\, (0.000993)$,
$\hat \sigma_1^2 = 0.0172 \,\, (0.00168)$,
$\hat \sigma_2^2 = 0.00911 \,\, (0.00105)$,
$\hat A = 0.264 \,\, (0.0294)$ and
$\hat \alpha = 1.506 \,\, (0.0773)$. The estimates for the renewal
main-shock arrival process $\hat\kappa$ and $\hat\beta$ implies a
mean waiting time of 23.11 days (SE: 1.023) between main-shocks with a standard
deviation of 25.09 days (SE: 1.291).  The shape parameter $\kappa$ has a 95\% 
confidence interval $(0.761, 0.936)$ indicating it deviates 
significantly from unity which further suggests that the classical \etas\  
may not be sufficient to model the heavy temporal clustering of main-shocks. 
The estimates $\hat A$ and $\hat\alpha$ 
imply that the expected number of aftershocks induced by a size 5,
5.5, 6, 6.5 and 7 magnitude earthquake are 0.26, 0.56, 1.19, 2.52 and
5.35 earthquakes respectively with a productivity of 0.534. 

\begin{figure}[b!]
	\centering
	\includegraphics[width=1\linewidth]{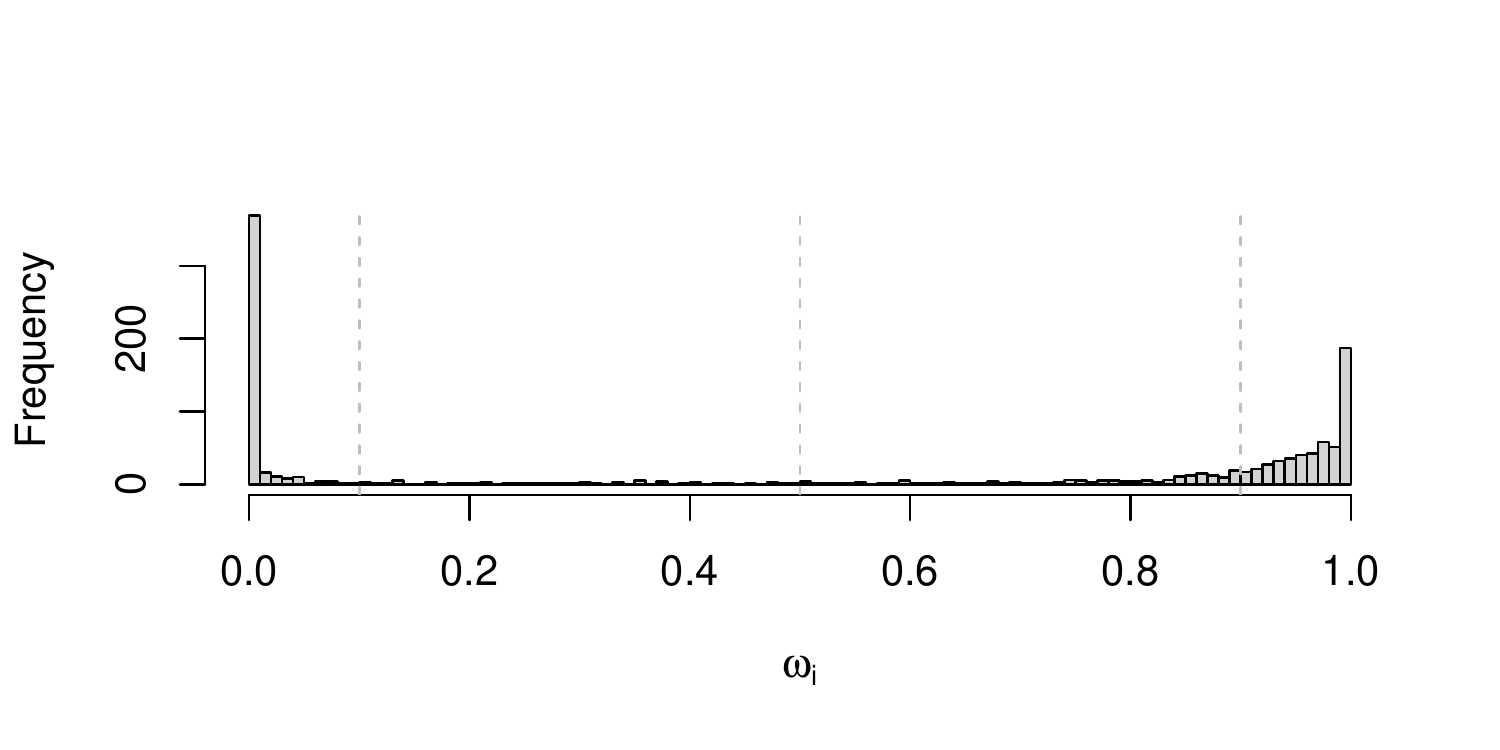}
	\caption{ \label{fig:as-probs} Frequency histogram of the
		main-shock probabilities $\omega_i$  implied
		by the \retas\ model for the NZ catalog with the
		amount of smoothing $\zeta = 0.5$. }
\end{figure}

Useful inferences can be drawn about the earthquake catalog 
based on the declustered main and aftershock probabilities. Since 
most of main-shock probabilities are close to either zero or one as seen in
Figure~\ref{fig:as-probs}, we can be reasonably confident in their
main-shock classification. For instance, employing a threshold 
probability of $0.5$, there are $684$ $(58.31\%)$ earthquakes 
in the catalog that would be classified as a main-shock, or if 
we employ the most-probable classification (as we did in the numerical 
experiments) then there are $703$ $(59.93\%)$ main-shocks.  In the
work of \cite{Harte2013} main-shocks represented approximately $35\%$
of the catalog which is significantly lower than predicted by the
\retas\ model. One possible reason for this major disparity is that
\citeauthor{Harte2013} only examines shallow earthquakes (less than
40km deep) while our work deals with large magnitude earthquakes.

\begin{figure}[t!]
	\centering
	\includegraphics[width=1\linewidth]{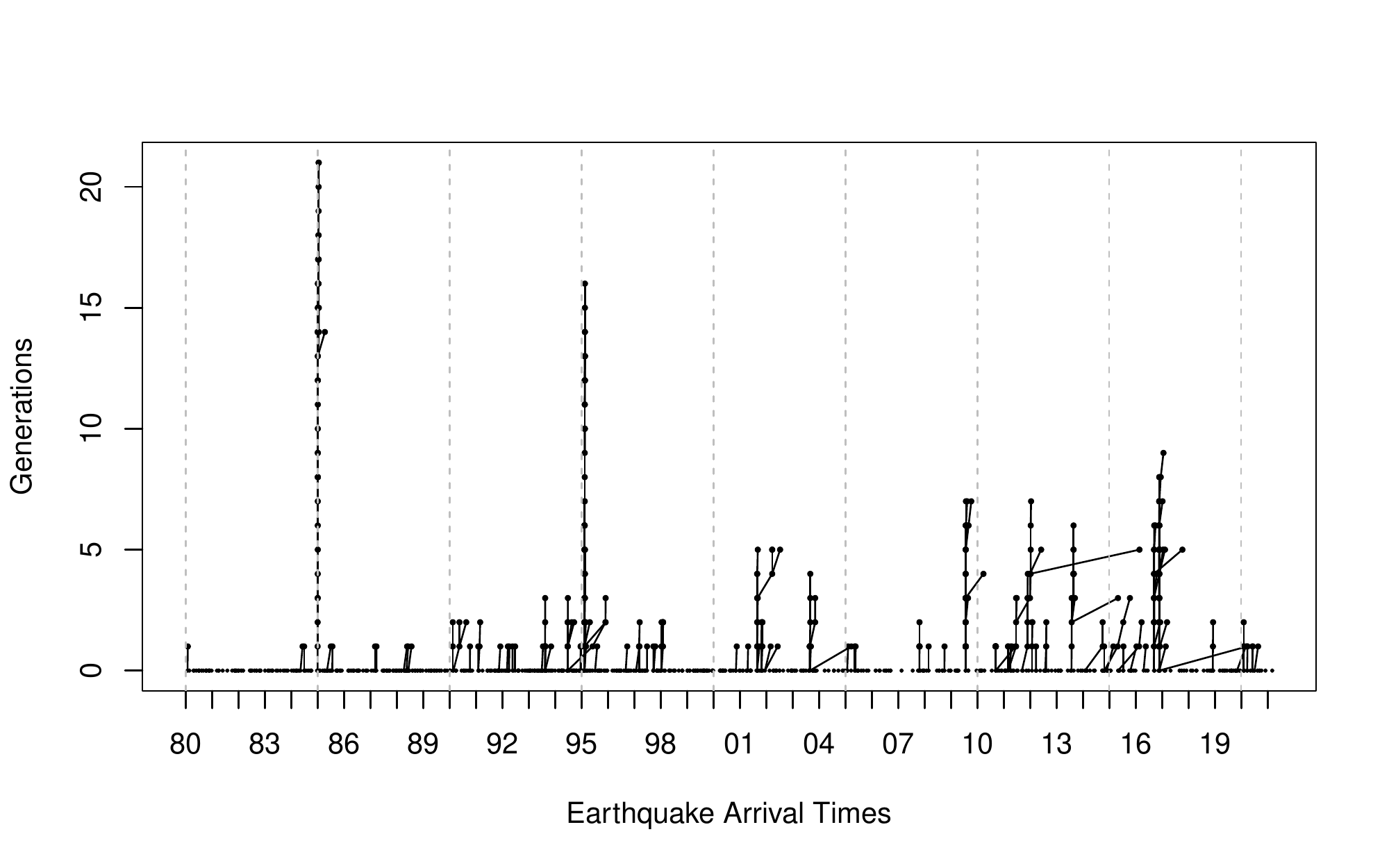}
	\caption{ \label{fig:as-probs1} Multi-generation branching 
		structure implied
		by the \retas\ model for the NZ catalog with optimal
		amount of smoothing $\zeta = 0.5$. }
\end{figure}

\begin{figure}[htp!]
	\centering
	\includegraphics[width=1\textwidth]{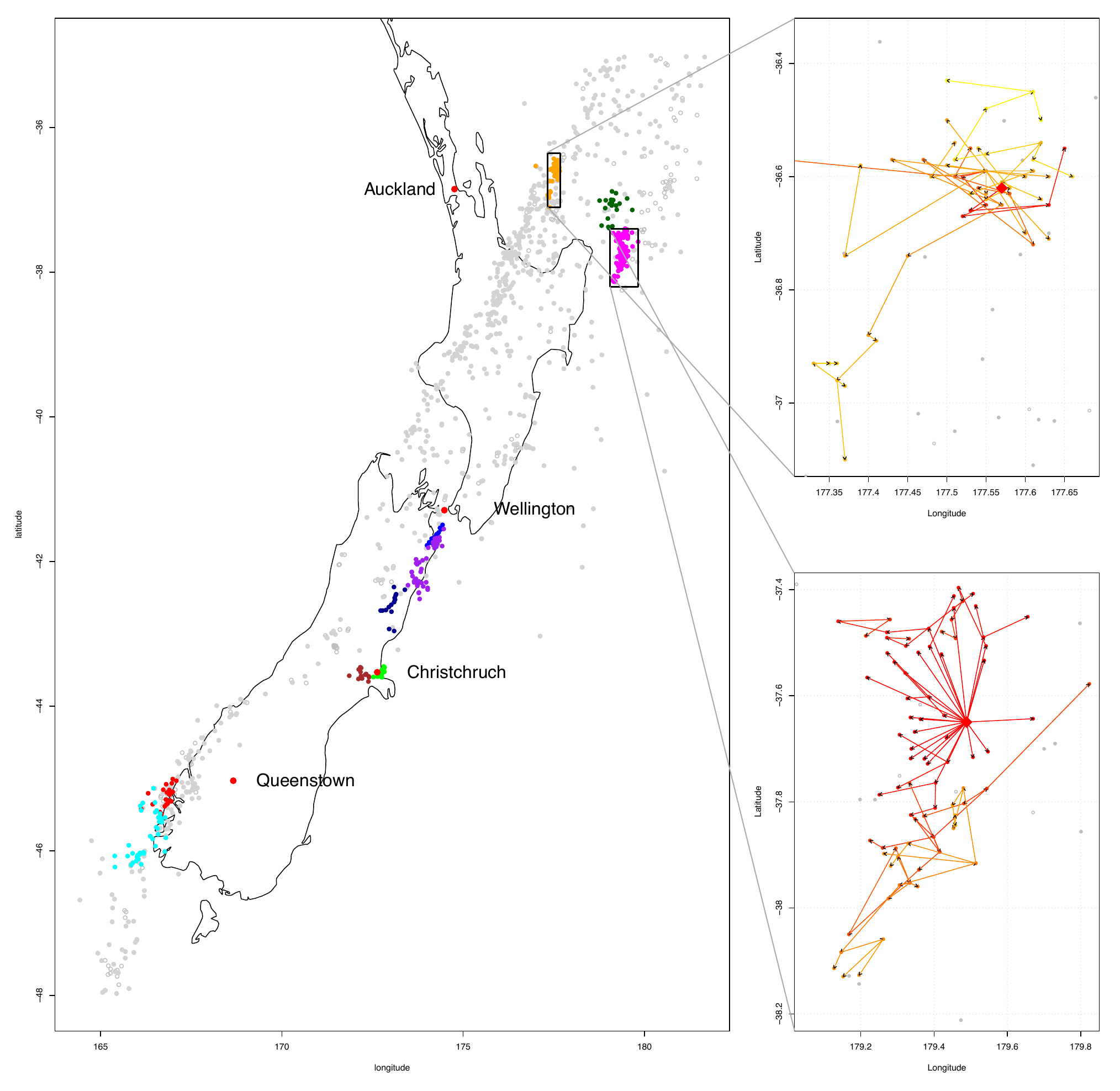}
	\caption{\label{fig:nzplot-branch2} Left plot contains main-shock (solid circles) and 
		aftershocks (open circles) based on fitted \retas\ model. Several large clusters are 
		represented by different colours on the plot. The two figures on the right
		are two large multi-generational aftershock sequences, in
		which the darker (lighter) shade arrows represent earlier (later) generation of aftershock.}
\end{figure}

The multi-generational structure implied by the fitted \retas\ model 
is shown in Figure~\ref{fig:as-probs1}. From the $703$ main-shocks (using the most-probable classification),
there were $73$ of these main-shocks which induced a first generation
aftershock with a total of $180$ aftershocks among them.  There was one
main-shock which induced $25$ generation one aftershocks of its own, and we
discuss this earthquake cluster next, as it is the most
destructive earthquakes in the catalog.  There were $46$ generation 
one aftershocks that induced a second generation, and 26 of these
generated a third generation aftershock. There are several large
clusters with many generations of aftershocks in the catalog. 
The largest cluster 
in the catalog begun on February 5th, 1995 at 22:51:02.31. It was 
located on the north-east region of New-Zealand 
($171.49^\circ$~N and $37.65^\circ$~S, lower right plot in Figure~\ref{fig:nzplot-branch2}) with a magnitude of 7.15 
(being the fourth largest magnitude earthquake in the catalog). The declustering algorithm implies that this earthquake 
induced 25 aftershocks directly from it. 
There are a total of sixteen generations of aftershock that have resulted from this devastating  earthquake. The size of the cluster (including the main-shock) 
is $84$ earthquakes with the second generation having $14$ earthquakes and the 
sixth generation having $7$ earthquakes. 

The second largest cluster with $53$ aftershocks over $21$ generations (cf. 
upper right plot in Figure~\ref{fig:nzplot-branch2}) originated on 
December 18th, 1984 at 11:16:58.12 at 177.57 $^\circ E$ 
and 36.63 $^\circ S$ with a main-shock magnitude of 5.24. Most aftershocks 
in this sequence only had one, two or three 
directly induced aftershocks, except for the earthquake that occurred 
two days later on December 30th, 1984 at 21:36:54 at 
177.55 $^\circ E$ and 36.59 $^\circ S$ with magnitude $6.70$, which 
was a $13$-th generation of aftershock in the cluster, and 
induced directly $11$ aftershocks of its own. The third largest cluster had 
ten generations with a cluster size of $52$ and occurred in North Canterbury
on November 13th, 2016 (see purple cluster in Figure~\ref{fig:nzplot-branch2}). 

The largest magnitude earthquake occurred on November 13th, 2010 
at 11:02:56 at $173.02^\circ$~N and $42.69^\circ$~S with a 
magnitude of $7.82$ (dark-blue cluster). It had five generations of 
aftershock with 15 aftershocks. The main-shock induced directly ten 
aftershocks on its own. The second largest magnitude at 
$7.80$ has seven generations of aftershock with a total cluster size
of 40 (cyan cluster). In contrast, the third largest magnitude at $7.20$ only
had one generation of aftershock with thirteen aftershocks 
(brown cluster).

Figure~\ref{fig:nzplot-branch2} shows several clusters 
that have been identified by the \retas\ model with different
colours representing the different clusters. The solid circles 
indicate main-shocks while the open circles indicate 
aftershocks. Some of the large clusters includes; Bay of Plenty in 
1984-1985 (orange), East Cape in February 1995 (magenta), 
the Fiordland earthquake in August 2003 (red), Dusky Sound
in July 2009 (cyan), Darfield in September 2010 known as 
the 2010 Canterbury earthquake (brown), Christchurch in February
2011 (green), Cook Strait in February 2011 which includes 
the 2013 Seddon earthquake (blue), NE of east Cape in September 
2016 called the Te Araroa earthquake (dark-green), and two 
clusters in North Canterbury in November 2011 in which 
one contains the 2016 Kaikoura earthquake (dark-blue) and 
a second cluster (purple). 

\begin{appendix}
	\section{Derivation of smoothed most recent main-shock probabilities $q_{ij}$}\label{appA}
	For compactness of notation, in the following, we define
	$z_i := (t_i,x_i,y_i,m_i)$ as the time, coordinates and magnitude
	of the $i$th earthquake. The backward recursion smooths the filtered
	probabilities $p_{ij} = P(I(t_i) = j | \mathcal{H}_{t_i-})$ calculated during likelihood evaluation based on weights derived from the conditional
	density
	$p(z_{i:n}, t_{n+1} > T | \mathcal{H}_{t_i-}, I(t_i) = j)$
	which follows from Bayes' theorem
	\begin{align}
	\label{eq:qij-app}
	q_{ij} 
	& =  \frac{p(z_{i:n}, t_{n+1} > T \mid  z_{1:i-1}, I(t_i) = j)}{ \sum_{k=1}^{i-1} p\brackets{z_{i:n}, t_{n+1} > T  \mid z_{1:i-1}, I(t_i) = k} \, p_{ik}} \, p_{ij},
	\end{align}
	but its evaluation is not directly possible. For the purpose of evaluating~\eqref{eq:qij-app}, we denote $f_{ij}$ as the joint density of $z_{i:n}$ and $t_{n+1} > T$ conditional on $z_{1:i-1}$ and $I(t_i) = j$, that is, $f_{ij} = p(z_{i:n}, t_{n+1} > T | z_{1:i-1}, I(t_i) = j)$. Then there exists a relationship between 
	the values $f_{ij}$ and $f_{i+1,j}$ by again applying Bayes' theorem
	\begin{align*}
	f_{ij} 
	& = p(z_i | z_{1:i-1}, I(t_i) = j) \,
	p(z_{i+1:n}, t_{n+1} > T \mid z_{1:i}, I(t_i) = j) \notag \\
	& = \psi_{ij} \, \sum_{k=1}^{i} p(z_{i+1:n}, t_{n+1} > T \mid z_{1:i}, I(t_i) = j, I(t_{i+1}) = k) 
	\, \P \brackets{I(t_{i+1}) = k \mid z_{1:i}, I(t_i) = j} 
	\end{align*}
	where 
	$$\psi_{ij} = p(z_i \mid z_{1:i-1}, I(t_i) = j)= (\mu(t_i - t_j)\nu(x_i,y_i) + \phi(t_i)) 
	e^ {- \int_{t_{i-1}}^{t_{i}} \mu(s)\rmd s - \{\Phi(t_i) - \Phi(t_{i-1})\} },
	$$ is the conditional density of $z_i$ conditioned on $z_{1:i-1}$ and $I(t_i) = j$.
	
	However, since the condition contains both $I(t_i) = j$ and
	$I(t_{i+1}) = k$, the only terms that contributes to the summation
	are when $k \in \set{i,j}$.  Furthermore, since the main-shock renewal
	time at $t_{i+1}$ is included in the condition, the additional
	information $I(t_i) = j$ in the condition is irrelevant as only the
	most recent renewal time provides information about the future
	evolution of the process conditional on the history. Therefore,
	calculation of $f_{ij}$ reduces to the following
	\begin{align}
	\label{eq:fij1-app}
	f_{ij} & = \psi_{ij} \, \sum_{k \in \set{i,j}} p(z_{i+1:n}, t_{n+1} > T \mid z_{1:i}, I(t_{i+1}) = k) 
	\, \P \brackets{I(t_{i+1}) = k \mid z_{1:i}, I(t_i) = j} \notag \\
	& =  \psi_{ij} \, \sum_{k \in \set{i,j}} f_{i+1,k}
	\, \P \brackets{I(t_{i+1}) = k \mid z_{1:i}, I(t_i) = j},
	\end{align}
	which depends on $f_{i+1,j}$. Since the most recent main-shock indicator function $I(t_{i+1})$ has two possible values  $I(t_i)$ or $i$ which depends on $B_i = 0$ (main-shock) or $B_i = j$, $ j < i$ (aftershock), $f_{ij}$ simplifies to the following
	\begin{align}
	f_{ij} & =  \psi_{ij} \bigg[f_{i+1,j}
	\, \P \brackets{I(t_{i+1}) = j \mid z_{1:i}, I(t_i) = j} + f_{i+1,i}
	\, \P \brackets{I(t_{i+1}) = i \mid z_{1:i}, I(t_i) = j}\bigg]\notag \\
	& =  \psi_{ij} \bigg[f_{i+1,j}
	\, \P \brackets{B_i \neq 0 \mid z_{1:i}, I(t_i) = j} + f_{i+1,i}
	\, \P \brackets{B_i = 0 \mid z_{1:i}, I(t_i) = j}\bigg]\notag \\
	& = \psi_{ij} \bigg[f_{i+1,j}
	\, \frac{\phi(t_i,x_i,y_i,m_i)}{\mu(t_i - t_j)\nu(x_i,y_i) + \phi(t_i,x_i,y_i,m_i)} + f_{i+1,i}
	\, 	\frac{\mu(t_i - t_j)\nu(x_i,y_i)}{\mu(t_i - t_j)\nu(x_i,y_i) + \phi(t_i,x_i,y_i,m_i)} \bigg] \notag \\ \label{eq:fij-rec-app}
	& = e^{-\int_{t_{i-1}}^{t_{i}} \mu(s-t_j)\rmd s  -(\Phi(t_{i})-\Phi(t_{i-1}))} \bigg[f_{i+1,j}
	\, \phi(t_i,x_i,y_i,m_i)+ f_{i+1,i}
	\, 	\mu(t_i - t_j)\nu(x_i,y_i)\bigg], 
	\end{align}
	which depends on future values of $f_{i+1,j}$ and a backward recursion exists with starting condition 
	\begin{equation*}
	f_{n+1,j} = P(t_{n+1} > T | z_{1:n}, t_{n+1} > T, I(t_{n+1}) = j) = e^{-\int_{t_{n}}^{T} \mu(s-t_j)\rmd s - \{\Phi(T) - \Phi(t_n)\}} 
	\end{equation*}
	for $j = 1,\dotsc,n$. 
	Hence, the smoothed most recent main-shock probabilities are given by
	\begin{equation}
	q_{ij} = \frac{f_{ij}}{\sum_{k=1}^{i-1} f_{ik} p_{ik}} \, p_{ij},
	\end{equation}
	for $ i = n,\dotsc, 2$ and $j = 1,\dotsc,i-1.$
	
	\section{Derivation of smoothed branching structure probabilities $\omega_{i}$ and $\pi_{ij}$ \label{appB}}
	First, by applying Bayes' theorem, the smoothed main-shock probabilities $\omega_{ij}$ are given by
	\begin{align}
	\omega_{ij} & = \P (B_i = 0, I(t_i) = j | z_{1:n}, t_{n+1} > T) \notag \\
	& =  \P \brackets{B_i = 0 | z_{1:n}, t_{n+1} > T, I(t_i) = j} \, 
	\P \brackets{I(t_i) = j | z_{1:n}, z_{n+1} > T } \notag \\
	& = \frac{p(z_{i+1:n}, t_{n+1} > T | z_{1:i}, B_i = 0, I(t_i) = j) \, 
		\P\brackets{M_i = 0 | z_{1:i}, I(t_i) = j}}{p(z_{i+1:n}, t_{n+1} > T | z_{1:i}, I(t_i) = j) } \, q_{ij} \notag \\
	& = \frac{p(z_{i+1:n}, t_{n+1} > T | z_{1:i},  I(t_{i+1}) = i) \, 
		\P\brackets{B_i = 0 | z_{1:i}, I(t_i) = j}}{p(z_{i+1:n}, t_{n+1} > T | z_{1:i}, I(t_i) = j) } \, q_{ij}.
	\end{align}
	Then applying a similar simplification to that used in~\eqref{eq:fij-rec-app} and the definition of $f_{ij}$ the following holds
	\begin{equation}
	\omega_{ij} = \frac{f_{i+1,i} \, \mu(t_i - t_j)\nu(x_i,y_i) }{f_{i+1,j} \, \phi(t_i,x_i,y_i,m_i) + f_{i+1,i} \, \mu(t_i - t_j)\nu(x_i,y_i)} \, q_{ij},
	\label{eq:omega-rec-app}
	\end{equation}
	for which the smoothed main-shock probability is obtained by summing over all possible indices 
	\begin{equation}
	\omega_{i} = \sum_{j=1}^{i-1} \omega_{ij}.
	\label{eq:omega-i-app}
	\end{equation}
	
	Calculation of the smoothed aftershock probabilities $\pi_{ij}$ are calculated by again applying Bayes' theorem and take the form
	\begin{align*}
	\pi_{ij} & = \P \brackets{B_i = j | z_{1:n}, t_{n+1} > T} \\
	& = \sum_{k=1}^{i-1} \P \brackets{B_i = j | z_{1:n}, t_{n+1} > T, I(t_i) = k} \, q_{ik} \\
	& = \sum_{k=1}^{i-1} \frac{p(z_{i+1:n}, t_{n+1} > T | z_{1:i}, I(t_i) = k, B_i = j) \, 
		\P \brackets{B_i = j | z_{1:i}, I(t_i) = k}}{p(z_{i+1:n}, t_{n+1} > T \mid 
		z_{1:i}, I(t_i) = k)} \, q_{ik}
	\end{align*}
	The denominator in the above expression has already been calculated in~\eqref{eq:fij1-app} and implies the following expression to calculate $\pi_{ij}$
	\begin{align}
	\pi_{ij} & = \sum_{k=1}^{i-1} \frac{p(z_{i+1:n}, t_{n+1} > T \mid z_{1:i}, I(t_{i+1}) = k) \, 
		\frac{ k(m_i)  g(t_i - t_j) f(x_i - x_j,y_i - y_j)}{\mu(t_i - t_k)\nu(x_i,y_i) + \phi(t_i,x_i,y_i,m_i)}}{f_{i+1,k} \, \frac{\phi(t_i,x_i,y_i,m_i)}{\mu(t_i - t_k)\nu(x_i,y_i) + \phi(t_i,x_i,y_i,m_i)} + f_{i+1,i} \, \frac{\mu(t_i - t_k)\nu(x_i,y_i)}{\mu(t_i - t_k)\nu(x_i,y_i) + \phi(t_i,x_i,y_i,m_i)}} \, q_{ik} \notag \\ 	\label{eq:pi-rec-app}
	& = \sum_{k=1}^{i-1} \frac{f_{i+1,k} \, k(m_i)  g(t_i - t_j) f(x_i - x_j,y_i - y_j)}{f_{i+1,k} \, \phi(t_i,x_i,y_i,m_i) + f_{i+1,i} \, \mu(t_i - t_k)\nu(x_i,y_i)} \, q_{ik} .
	\end{align}
\end{appendix}

\bibliographystyle{apalike}
\bibliography{RETAS-declustering-arXiv}

\end{document}